\documentclass[12pt]{article}
\usepackage{amsfonts,amssymb,amsmath}
\usepackage{graphics}
\usepackage{lscape}
\usepackage{makeidx}
\usepackage{theorem}

\newtheorem{theorem}{Proposition}
 
\newcommand{\eps}{{\varepsilon}}
\newcommand{\be}{\begin{equation}}
\newcommand{\ee}{\end{equation}}
\newcommand{\bea}{\begin{eqnarray}}
\newcommand{\eea}{\end{eqnarray}}

\newcommand{\ra}{\rightarrow}

\newcommand{\vr}{\varepsilon}

\begin{document}

\newpage
\pagestyle{empty}

\vfill

\rightline{DSF-12/05}

\vfill

\begin{center}

    {\Large \textbf{\textsf{Standard and Non-standard Extensions of Lie algebras}}}

    \vspace{10mm}

    {\large L. A. Forte, A. Sciarrino}

    \vspace{10mm}

    \emph{  Dipartimento di Scienze Fisiche, Universit{\`a} di Napoli
    ``Federico II''}

    \emph{and I.N.F.N., Sezione di Napoli,}

    \emph{Complesso Universitario di Monte S. Angelo,}

    \emph{Via Cinthia, I-80126 Naples, Italy}

    \vspace{7mm}
\begin{center}
 {\bf \large Revised version}
 \end{center}

    \vspace{12mm}

    \end{center}

    \vspace{7mm}

\begin{abstract}
We study the problem of quadruple extensions of simple Lie
algebras. We find that, adding a new simple root $\alpha_{+4}$, it
is not possible to have an extended Kac-Moody algebra described by
a Dynkin-Kac diagram with simple links and no loops between the dots, while it is
possible if $\alpha_{+4}$ is a Borcherds imaginary simple root. We
also comment on the root lattices of these new algebras. The
folding procedure is applied to the simply-laced triple extended
Lie algebras, obtaining all  the non-simply laced ones.  Non-
standard extension procedures for a class of Lie algebras are
proposed.  It is shown that the  $2$-extensions of  $E_{8}$, with
a dot simply linked  to the Dynkin-Kac diagram of  $E_{9}$, are
rank 10 subalgebras of $E_{10}$. Finally  the simple root systems
of a set of  rank 11 subalgebras of $E_{11}$,  containing as
sub-algebra $E_{10}$, are explicitly  written.
\end{abstract}

      \vfill

    \vfill
    \vspace*{3mm}

    \hrule

\vspace*{2mm} \emph{E-mail:} \texttt{forte@na.infn.it} \;\;\
\texttt{sciarrino@na.infn.it}
    \vspace*{3mm}

    \newpage
    \pagestyle{plain}
    \setcounter{page}{1}
    \baselineskip=16pt
    \parindent=0pt

\section{Introduction}

It has been conjectured  by Peter West \cite{West01}  that  the
still elusive M-Theory possesses a rank eleven Kac-Moody symmetry
algebra, called $E_{11}$, that is the triple extended or very
extended $E_{8}$ algebra. Very extended algebras can be defined
for any finite-dimensional Lie algebra ${\cal G}$ \cite{GOW}. So
it is tempting to argue that other theories, which are associated
with other triple extensions of Lie algebras, may exist. Indeed
the same analysis was applied to a conjectured extension of the
eleven dimensional supergravity \cite{SW01}, \cite{West04}. More
generally, it has been proposed that the closed bosonic  string in
D dimensions and type I supergravity and pure gravity theories
exhibit a Kac-Moody symmetry algebra, respectively identified as
the triple extensions of the $D$ and $A$ series \cite{LW},
\cite{SW04}, \cite{KSW04}. This conjecture is supported by
dimensional reduction and by the so called cosmological billiards
\cite{DHN}, \cite{DBNS}, \cite{EHTW}. Then it is natural to look
for a more general symmetry algebra which can include all these
Kac-Moody algebras as particular cases. So we address the question
of how to go beyond $\mathcal{G}^{+++}$ algebras; we find that the
adjoint of a new simple root $\alpha_{+4}$ introduces multiple
links and loops in the structure of the 4-extended algebra, if
$\alpha_{+4}$ is an ordinary Kac-Moody simple root, while the
"simple-links" structure is preserved if we allow $\alpha_{+4}$ is
a Borcherds (imaginary) simple root.

The (first) extension of a finite-dimensional Lie algebra is the
construction of (untwisted) affine Kac-Moody algebras, which are
obtained adding to the simple roots of any finite-dimensional Lie
algebra $\mathcal{G}$ a root $\alpha_{0}$ that is the opposite of
the highest root (h.r.) plus a light-like vector $k_{+}$, in order
to make $\alpha_{0}$ linearly independent from the system of the
simple roots of $\mathcal{G}$, keeping unchanged  its length, and
are denoted as $\mathcal{G^{+}}$ or  $\mathcal{\widehat{G}}$ or
$\mathcal{G}^{1}$. This procedure for the simply laced algebras of
the $D_{N}$ and $E_{N}$ series can be formulated as the addition
to the simple root system of $\mathcal{G}$ of another root
$\alpha_{0}$, that is the opposite of the unique fundamental
weight of length 2, which is in the root lattice of the algebra,
plus a light-like vector $k_{+}$. The light-like vector can be
considered to belong to a 2-dim. Lorentzian lattice, usually
denoted II$^{1,1}$, and the double extension  or overextension of
$\mathcal{G}$, denoted  $\mathcal{G^{++}}$, is
 obtained by adding a
new simple root, of length 2, which is formed by the sum of the
two light like vectors $k_{\pm}$, $(k_{+}, k_{-}) = 1$, spanning
II$^{1,1}$. The triple extended or very extended $\mathcal{G}$,
denoted  $\mathcal{G^{+++}}$, is obtained adding a new simple root
of length 2, which belongs to a new copy of the lattice
II$^{1,1}$, plus $k_{+}$. In this way, an indefinite Kac-Moody or
Lorentzian algebra of rank $r+3$ is obtained whose roots belong to
a Lorentzian lattice of dimension $r+4$, so it is natural to
wonder if an indefinite Kac-Moody algebra of rank $r+4$ can be
obtained by a further extension. Moreover, let us notice that in
the lattice II$^{1,1}$ vectors of negative length do exist (see
Appendix \ref{lattice}). From this remark one can make an
extension of $\mathcal{G}$ adding a new root, that is the opposite
of any  fundamental weight,  that can be written as  linear
combination with integer coefficients  of the simple  roots of the
algebra, plus a suitable  element of the lattice II$^{1,1}$ in
order to have an independent new simple root of length 2. This
construction will be discussed below. It has been pointed out in
\cite{FN} that the structure of subalgebras of hyperbolic
Kac-Moody, in general of $2$-extended (overextended) Lie algebras,
is very rich and surprising.  Some of the results of that paper
can be generalized to more general extensions and we comment on
this point below. This paper is organized as it follows. In Sec.
\ref{triple}, to make the paper self contained, we recall the
well-known construction of the 3-extended Lie algebras. We show
that the 4-extended algebras are described by Dynkin-Kac diagrams
with loops and multiple links (so their structure is quite
different from that of the 1-, 2- and 3-extended algebras). In
particular, we show that we can not have a situation in which the
new (fourth) simple root $\alpha_{+4}$ is simply linked to
$\alpha_{+3}$, unless we let $\alpha_{+4}$ be a Borcherds simple
root (with squared norm zero or negative). So we also study the
possibility to have a Borcherds extension of the
$\mathcal{G}^{+++}$ algebras, but this extension has sense only
when the algebra $\mathcal{G}$ is simply-laced (see Appendix
\ref{Borcherds} for the definition of a Borcherds algebra
\cite{Bor}). In Sec. \ref{folding}, we show that all the
non-simply laced 3-extended Lie algebras can be obtained by
\emph{folding} the simply-laced ones. In Sec. \ref{nonstandard},
we discuss non standard extension procedures, discussing in detail
a few examples which may be of physical interest. In Sec.
\ref{subalgebras},  we show that  the algebras obtained by some
general non standard procedure, but not for all the procedures,
are indeed subalgebras, of the same rank, of the standard triple
extended algebras.  In particular  we prove that any non standard
$1$-extension of $E_{9}$, with a root simply linked to a simple
roots of $E_{9}$, is a subalgebra of  $E_{10}$. The simple root
systems of a set of rank 11 subalgebras of $E_{11}$, containing as
subalgebra $E_{10}$, are explicitly written. Finally we present a
few conclusions and perspectives. To make the paper self-contained
two very short Appendices are added to recall the main features of
the $2$-dim Lorentzian lattice II$^{1,1}$ and of Borcherds
algebras.

\section{On extensions of $\mathcal{G}^{+++}$ algebras} \label{triple}

An excellent discussion of the mechanism of standard extensions of
Lie algebras can be found in \cite{GOW}, here we briefly recall
the essential points, mainly  to introduce the notation and to
make the paper self-consistent \footnote{Let us remember that the
standard extension for G$_{2}$ does not work, because G$_{2}$ is
the only finite Lie algebra for which it is not possible to
normalize the highest root $\theta$ such that $(\theta,\theta)=2$.
However, it's possible to extend G$_{2}$ with the following choice
of the extended roots: $\alpha_{+1}=k_{+} - \theta$, $\alpha_{+2}
= -(k_{+}+3k_{-})$ and $\alpha_{+3}=-(l_{+}+3l_{-}) + k_{+}$.
Anyway, G$_{2}^{+++}$ can be obtained by folding D$_{4}^{+++}$
(see Section \ref{folding}).}. Let $\mathcal{G}$ a simple Lie
algebra of rank $r$, with simple root system $\alpha_i$ ($i = 1,
\ldots, r$)  and root lattice $\Lambda_{\mathcal{G}} =
\bigoplus_{i=1}^{r}\mathbb{Z}\,\alpha_{i}$ (for the roots and
fundamental weights we use the notation of \cite{FSS}). Let us
consider also two copies of the lattice II$^{1,1}$, which we
indicate with II$^{1,1}_{k_{\pm}}$ and II$^{1,1}_{l_{\pm}}$. In
the following, we  consider indefinite Kac-Moody algebras with
root lattice included in the direct sum $\Lambda_{\mathcal{G}}\;
\oplus$ II$^{1,1}_{k_{\pm}} \oplus$ II$^{1,1}_{l_{\pm}}$.

Let $\mathcal{G^{+}}$ the extended Lie algebra (or affine
Kac-Moody algebra), with simple root system \{$\alpha_i$,
$\alpha_0 \equiv \alpha_{r+1} \equiv \alpha_{+1}= - h.r. + k_{+}\}
$ ($i = 1, \ldots, r$), where $h.r.$ denotes the highest root of
$\mathcal{G}$ (which is $\theta \equiv h.r. = \sum_{i=1}^{r}
a_{i}\alpha_{i}$ where $a_{i}$ are Kac marks, see \cite{FSS}), and
\be (k_{+}, k_{+})  = (k_{+}, \alpha_{i}) = 0. \ee
Let $\mathcal{G^{++}}$ the double extended or overextended Lie
algebra (actually a Lorentzian Kac-Moody algebra), with simple
root system \{$\alpha_j$, $\alpha_{r+2} \equiv \alpha_{+2}=
-(k_{+} + k_{-}) \}$ ($j = 1, \ldots, r+1$), where \be (k_{+},
k_{-})  = 1\, , \;\;\;\;\;\;\; (k_{-}, k_{-})  = (k_{-},
\alpha_{i}) = 0. \ee
Let's note that the root lattice of $\mathcal{G}^{+}$ is properly
contained in the direct sum $\Lambda_{\mathcal{G}}\; \oplus$
II$^{1,1}_{k_{\pm}}$, whereas the root lattice of
$\mathcal{G}^{++}$ coincides with the direct sum. To see this, it
is enough to obtain $k_{+}$ and $k_{-}$ from the other roots
$\alpha_{j}$ ($j = 1, \ldots, r+2$). In fact,
\begin{equation}
k_{+} = \theta + \alpha_{+1} = \sum_{i=1}^{r} a_{i}\,\alpha_{i} +
\alpha_{+1}
\end{equation}
\begin{equation}
k_{-} = - k_{+} - \alpha_{+2} = -\theta - \alpha_{+1} -
\alpha_{+2}
\end{equation}
so $k_{+}$ and $k_{-}$ are linear combinations (with integer
coefficients) of the $r+2$ simple roots of the $\mathcal{G}^{++}$
algebra. This means that with a change of basis (in the lattice),
we can pass from $\{\alpha_{j}\, (j = 1, \ldots, r+2)\}$ to
$\{\alpha_{i}\, (i = 1, \ldots, r);\, k_{+},\, k_{-}\}$, that is
the lattice $\Lambda_{\mathcal{G}} \oplus \mathbb{Z}\,\alpha_{+1}
\oplus \mathbb{Z}\,\alpha_{+2}$ coincides with
$\Lambda_{\mathcal{G}}\; \oplus$ II$^{1,1}_{k_{\pm}}$.

Now, let $\mathcal{G^{+++}}$ the  triple extended Lie algebra
(still a Lorentzian Kac-Moody algebra), with simple root system
\{$\alpha_k$, $\alpha_{r+3} \equiv \alpha_{+3}=   -(l_{+} + l_{-})
+ k_{+}$\} ($k = 1, \ldots, r+2$), where \be (\alpha_{i}, l_{\pm})
= (k_{\pm}, l_{\pm}) = (k_{\mp}, l_{\pm}) = (l_{\pm}, l_{\pm}) =
0\, , \;\;\;\;\;\;
   (l_{\mp}, l_{\pm})  =   1.
\ee
In this way starting from the $r-$dim. Euclidean lattice
$\Lambda_{\mathcal{G}}$, we have build up a $(r+4)-$dim.
Lorentzian lattice ${\bf \Lambda} \equiv \Lambda_{\mathcal{G}}\,
 \oplus $ II$^{1,1}_{k_{\pm}} \oplus$ II$^{1,1}_{l_{\pm}}$, with
signature $+ \ldots + - -$ ($r+2$ plus signs). The simple root
system of $\mathcal{G}^{+++}$ clearly spans a 3-dim. sub-lattice
of ${\bf \Lambda}$ (because $\alpha_{+3}$ only takes the direction
$l_{+} + l_{-}$ in the second lattice II$^{1,1}$, so the
orthogonal direction is lacking), so it is natural to wonder if it
is possible to extend further the $\mathcal{G}^{+++}$ algebra and
\textit{fill in} ${\bf \Lambda}$.
\newline
Motivated by the previous steps, one would just add another node
at the Dynkin diagram of a $\mathcal{G}^{+++}$ algebra, with a
simple link to the root $\alpha_{+3}$; in this way, as it happens
in the case of the $\mathcal{G}^{++}$ algebra, one would expect
that this construction fills the root lattice
$\Lambda_{\mathcal{G}}\, \oplus $II$^{1,1}_{k_{\pm}}\, \oplus\,
$II$^{1,1}_{l_{\pm}}$. In the following, we show that actually
this is not the case \footnote{Our discussion does not take account
for the introduction of another lattice II$^{1,1}$, which solves
the question only partially, because it moves the problem to fill
this new lattice.}.

As we already mentioned, the simple root $\alpha_{+3}$ contains
only the combination $l_{+} + l_{-}$, so if we want to span
completely also the second lattice II$^{1,1}$, we need a new
simple root $\alpha_{+4}$ which allows us to obtain $l_{+}$ and
$l_{-}$ separately. In this way, we are guaranteed that the root
lattice of the new algebra contains also all vectors which are
integer multiples of $l_{+}$ or $l_{-}\,$, so this lattice
coincides with ${\bf \Lambda}$. In the following, let's choose for
$\alpha_{+4}$ the more general form:
\begin{equation}
\alpha_{+4} \equiv a\, l_{+} + b\, l_{-} + c\, k_{+} + d\, k_{-}
\end{equation}
which allows us to make many considerations about the possible
extensions of the $\mathcal{G}^{+++}$ algebras \footnote{As it
happens also for $\alpha_{+2}$ and $\alpha_{+3}$, here we do not
consider the situation in which $\alpha_{+4}$ is also linked with
the simple roots $\alpha_{i}$ of $\mathcal{G}$. However  this simple choice
does not really change the conclusions of our analysis .}. First of all, let's obtain
the expression for $l_{+}$ and $l_{-}$ using the definition of
$\alpha_{+3}$ and $\alpha_{+4}$:
\begin{equation}
(a-b)\, l_{+} = (d-b-c)\, \theta + (d-b-c)\,\alpha_{+1} + d\,
\alpha_{+2} + b\, \alpha_{+3} + \alpha_{+4}\;,
\end{equation}
\begin{equation}
(b-a)\, l_{-} = (d-a-c)\, \theta + (d-a-c)\,\alpha_{+1} + d\,
\alpha_{+2} + a\, \alpha_{+3} + \alpha_{+4}.
\end{equation}
We see then that it is possible to obtain $l_{+}$ and $l_{-}$ as
linear combinations (over $\mathbb{Z}$) of the simple roots if and
only if $|\,a-b\,|=1$. Only in this case we are guaranteed that we
actually catch all the vectors in the lattice
II$^{1,1}_{l_{\pm}}$. With this in mind, let's find the other
relations $a, b, c, d$ have to satisfy, that is let us compute the
norm of $\alpha_{+4}$ and the scalar products with the other
simple roots (here we write only the relevant elements $a_{ij}$ of
the generalized Cartan matrix):
\begin{equation}
(\alpha_{+4})^{2} = 2\, (a\,b + c\,d)\;,
\end{equation}
\begin{equation}
a_{+4,\,+1} \equiv
2\,\frac{(\alpha_{+4},\alpha_{+1})}{(\alpha_{+4},\alpha_{+4})} =
\frac{d}{a\,b + c\,d}\;,\quad\quad a_{+4,\,+2} \equiv
2\,\frac{(\alpha_{+4},\alpha_{+2})}{(\alpha_{+4},\alpha_{+4})} =
\frac{-c-d}{a\,b + c\,d}\;,
\end{equation}
\begin{equation}
a_{+4,\,+3} \equiv
2\,\frac{(\alpha_{+4},\alpha_{+3})}{(\alpha_{+4},\alpha_{+4})} =
\frac{d-a-b}{a\,b + c\,d}\;.
\end{equation}
Let us limit ourselves to consider the case $(\alpha_{+4})^{2} = 2$
(to look for a "natural extension") and let all the four
coefficients be different from zero; we must have:
\begin{equation}
a\,b + c\,d = 1\;, \label{norm4}
\end{equation}
\begin{equation} \label{constraints}
d < 0\;,\quad\quad c \geq -d >0\;,\quad\quad d \leq a+b\;,
\end{equation}
because only with these constraints $\alpha_{+4}$ is an acceptable
simple root ({\`a} la Kac-Moody). In particular, the product $c\,
d$ must be negative, and as eq. (\ref{norm4}) holds, the product
$a\, b$ must be positive (actually, it must be at least 2), that
is $a$ and $b$ must have the same sign. A simple solution
\footnote{We thank C. Helfgott for pointing us this solution.} to
eqs. (\ref{norm4})-(\ref{constraints}) is the following:
\begin{equation}
a=2,\quad b=1,\quad c=1,\quad d=-1,
\end{equation}
where $\alpha_{+4}$ has norm 2 and the scalar products are
$(\alpha_{+4},\alpha_{+1})=-1$, $(\alpha_{+4},\alpha_{+2})=0$ and
$(\alpha_{+4},\alpha_{+3})=-4$. Furthermore, this solution
verifies also the condition $a-b=1$, so its root lattice is
precisely ${\bf \Lambda}$, but it presents one loop and a multiple
link with the (last) root $\alpha_{+3}$, so its structure is very
different from the other previous extensions \footnote{We thank A.
Kleinschmidt for suggesting us another solution $\alpha_{+4} =
\alpha + 3\,k_{+} - 2\,k_{-} -2\,l_{+}-3\,l_{-}$ in which
$\alpha_{+4}$ is also linked to the only simple root $\alpha$ of
su(2); this solution too presents loops and multiple links.}. This
situation is common to all solutions of eqs.
(\ref{norm4})-(\ref{constraints}): it is not possible (we stress:
without adding any II$^{1,1}$ or different lattice) to have a
simple link with $\alpha_{+4}$ and no loops (for example the
equation $d-a-b=-1$ is never satisfied if $a$ and $b$ are both
positive, because d is at least -1, and goes in contradiction with
the other relations if $a$ and $b$ are both negative). In this
sense, the procedure of standard extension stops at the third step
($\mathcal{G}^{+++}$).
\newline
It is clear that there exist infinite solutions to eqs.
(\ref{norm4})-(\ref{constraints}) (for example, it is enough to
take $a, b$ both positive, $d=-1$ and an opportune value of $c$ to
satisfy eq. (\ref{norm4}) and the second of eqs.
(\ref{constraints})), but only those with $|a-b|=1$ have the
property that their root lattice coincides with ${\bf \Lambda}$.
We can also try (in $\alpha_{+4}$) to put one coefficient equal
to zero and to explore more specific cases. An investigation, case by
case, shows that:

\begin{itemize}
    \item $d=0$ : the scalar products imply $ab=1$ and $a+b > 0$, that is
    $a=b=1$, but the case $a=b$ is not acceptable because, in this case, 
    $\alpha_{+4}$ is a linear combination of the other simple
    roots;
    \item $c=0$ : the scalar products with $\alpha_{+1}$ and
    $\alpha_{+2}$ are respectively proportional to $d$ and $-d$,
    so one of the two is positive and has the wrong sign, or $d=0$
    and then $\alpha_{+4}$ is a linear combination of the other
    simple roots;
    \item $a=0$ : this case implies $c=-1$, but then $d=-1$ too
    and the scalar product with $\alpha_{+2}$ is positive;
    \item $b=0$ : it has the same problem as the case $a=0$.
\end{itemize}
If we put equal to zero two coefficients, the only possibility is
to have $c=d=0$ (we can not take equal to zero a coefficient of
$l_{+}$ or $l_{-}$ and a coefficient of $k_{+}$ or $k_{-}$,
because  $\alpha_{+4}$ would have vanishing norm),  but in this
case, as already discussed, we have problems.

So far, we have seen that it is never possible to add a
(Kac-Moody) simple root in order to have a simple link with
$\alpha_{+3}$; besides this, the possible solutions which span the
whole lattice ${\bf \Lambda}$ are restricted to the condition
$|a-b|=1$. Yet, if we abandon the condition that $\alpha_{+4}$ is
a Kac-Moody simple root and let it be a Borcherds (imaginary)
simple root, the situation changes. In fact, we can easily find an
imaginary root of the kind:
\begin{equation}
\alpha_{+4}= a\, l_{+} + b\, l_{-}
\end{equation}
which has scalar product equal to $-1$ with $\alpha_{+3}$: it is
only necessary that $b=1-a$, that is $\alpha_{+4}= a\, l_{+} +
(1-a)\, l_{-}$, with norm $2\,a\,(1-a)$. Then if $a=0$ or $a=1$,
$\alpha_{+4}$ has norm zero, in all the other cases its norm is
negative, that is $\alpha_{+4}$ is a good Borcherds simple root.
If we want to fill in ${\bf \Lambda}$ with the introduction of a
Borcherds simple root, we have always to fulfill the condition
$|a-b|=1$ (this condition is independent from the norm of
$\alpha_{+4}$), but now it is possible to have both
$(\alpha_{+4},\alpha_{+3})= -1$ and, at the same time, to span the
whole lattice ${\bf \Lambda}$. In fact the values $a=1, b=0$ (and
viceversa $a=0, b=1$), which correspond to $\alpha_{+4}=l_{+}$
($\alpha_{+4}=l_{-}$), are the only ones which allow to have:
\begin{equation}
\Lambda_{roots}(\mathcal{BG\mathcal{^{+++}}}) \equiv
\Lambda_{roots}(\mathcal{G\mathcal{^{++}}}) \oplus
\mbox{II}^{1,1}_{l_{\pm}} = {\bf \Lambda}\;,
\end{equation}
\begin{equation}
(\alpha_{+4},\alpha_{+3}) = -1\,, \label{eq.-1}
\end{equation}
as opposite to the Kac-Moody case, where, as we already said, it's
never possible to satisfy eq. (\ref{eq.-1}) and only some
solutions allow to fill in ${\bf \Lambda}$ (we have called
$\mathcal{BG\mathcal{^{+++}}}$ the Borcherds extension of
$\mathcal{G\mathcal{^{+++}}}$ corresponding to $\alpha_{+4}=l_{+}$
or $\alpha_{+4}=l_{-}$).  Denoting by a crossed
dot the Borcherds simple root $\alpha_{+4}$, we can draw the
Dynkin diagram of this Borcherds algebra in the following way:

\bigskip
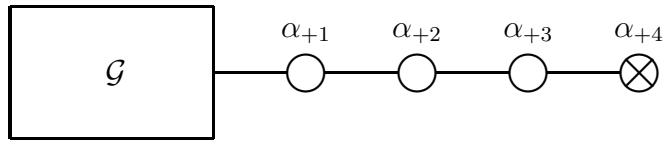
\begin{center}
    \begin{picture}(80,20) \thicklines
        \multiput(42,0)(42,0){4}{\circle{14}}
        \put(-30,0){\makebox(0,0){$\mathcal{G}$}}
        \put(42,15){\makebox(0,0){$\alpha_{+1}$}}
        \put(84,15){\makebox(0,0){$\alpha_{+2}$}}
        \put(126,15){\makebox(0,0){$\alpha_{+3}$}}
        \put(168,15){\makebox(0,0){$\alpha_{+4}$}}
        \put(7,0){\line(1,0){28}}
        \put(49,0){\line(1,0){28}}
        \put(91,0){\line(1,0){28}}
        \put(7,0){\line(0,-1){25}}
        \put(7,0){\line(0,1){25}}
        \put(7,25){\line(-1,0){77}}
        \put(7,-25){\line(-1,0){77}}
        \put(-70,0){\line(0,1){25}}
        \put(-70,0){\line(0,-1){25}}
        \put(133,0){\line(1,0){28}}
%        \put(133,0){\line(1,0){4}}\put(141,0){\line(1,0){4}}
%        \put(149,0){\line(1,0){4}}\put(157,0){\line(1,0){4}}
%        \put(132,3){\line(1,0){30}}
%        \put(132,-3){\line(1,0){30}}
        \put(173,-5){\line(-1,1){10}}
        \put(163,-5){\line(1,1){10}}
    \end{picture}
\end{center}
\bigskip
\bigskip
\bigskip

Actually, this construction makes sense if $\mathcal{G}$ is
simply-laced, because otherwise the Cartan matrix is not
well-defined. In fact, if $\alpha_{+4}$ is an imaginary
(isotropic) root, we cannot define the extended Cartan matrix as
$2\,\frac{(\alpha_{i},\alpha_{j})}{(\alpha_{i},\alpha_{i})}$ for
all $\alpha_{i}$ because $\alpha_{+4}^{2}=0$. We have the same
problem if we add an imaginary simple root with negative squared
norm (as previously recalled, in II$^{1,1}$ there are infinite
vectors whose squared norm is negative), because we shall have
positive elements out of the principal diagonal. The solution is
then to consider a $\mathcal{G}$ simply-laced and define the
extended Cartan matrix by the scalar products between all the
simple roots: $b_{i,j}:=(\alpha_{i},\alpha_{j})$. So, while the
extension of $\mathcal{G}$ is possible up to $\mathcal{G}^{+++}$
for any finite $\mathcal{G}$, in the case of a Borcherds extension
it is necessary to choose for $\mathcal{G}$ a simply-laced algebra
(in fact, a Borcherds algebra is defined only on a symmetric
Cartan matrix). To summarize our result, we have proven the
following {\theorem The extension of  $\mathcal{G}^{+++}$ algebra,
whose simple root system spans completely ${\bf \Lambda}$ and
whose Dynkin-Kac diagram has no loops and only simple links between the dots, is
the Borcherds algebra $\mathcal{BG\mathcal{^{+++}}}$ (with
$\mathcal{G}$ simply-laced).}

The particular solution $a=1, b=0$ (or viceversa) looks like the
same construction of $\mathcal{G}^{+}$ and $\mathcal{G}^{++}$, as
the role of $k_{\pm}$ is now played by $l_{\pm}$, with the
difference that $\alpha_{+1}$ contains also a root of a simple Lie
algebra, while $\alpha_{+4}$ doesn't. This observation suggests
that it's possible to fuse together two (finite) simple Lie
algebras, let's say $\mathcal{G}$ and $\mathcal{G}^{\prime}$,
adding the highest root $\theta^{\prime}$ of
$\mathcal{G}^{\prime}$ to $\alpha_{+4}$. In this way,
$\alpha_{+4}^{\prime} \equiv \alpha_{+4} - \theta^{\prime}$ is
again a Kac-Moody simple root, indeed the affine root of
$\mathcal{G}^{\prime\,+}$. Anyway, we don't insist on this point,
because there are many possible ways to fuse together two (or
more) finite dimensional simple Lie algebras (with or without the
introduction of intermediate Kac-Moody or Borcherds algebras).
\newline
\textit{Remark}: In this section, we have seen that triple
extended Lie algebras $\mathcal{G}^{+++}$ have their root lattice
properly included in ${\bf \Lambda} =
\Lambda_{\mathcal{G}}\,\oplus$ II$^{1,1} \oplus$ II$^{1,1}$ (in
particular, their root lattice is Lorentzian with just one
negative eigenvalue). Among the 4-extensions we considered in this Section,
 we have shown that there are many algebras (of
Kac-Moody or generalized Kac-Moody type) whose root lattice
coincides with ${\bf \Lambda}$, which is Lorentzian in a more
general sense (it has two negative eigenvalues). These seem to be
the first algebras obtained in literature with this kind of
lattice, together with some similar algebras studied by Harvey and
Moore in \cite{HM95}, \cite{HM96} (strictly speaking, their
algebras are not Borcherds algebras, because they could not
satisfy some grading conditions in the characterization of
generalized Kac-Moody algebras, while our Borcherds algebras are
\emph{true} Borcherds algebras because we have constructed them
from an acceptable generalized Cartan matrix). Our result seems to
be in contrast with the statement of \cite{Klein}, according to
which algebras whose root lattices are of the kind $\Gamma^{p,q}$
(that is, with a signature with more that one negative sign) cannot
be described in terms of generators and relations similar to
Kac-Moody or Borcherds algebras and belong to a new class of Lie
algebras. That result has been found looking for  Lie algebras of
the physical states of a vertex algebra constructed on a general
even self-dual lattice $\Gamma^{p,q}$; maybe the condition of
physical states is too strong and is not compatible with the
algebras constructed by us.

\section{Folding of triple extended Lie algebras}  \label{folding}

The  \emph{folding} technique is a simple and powerful method to
find class of singular subalgebras of finite Lie algebras as well
as of affine or indefinite Kac-Moody algebras. The starting point
is to use the symmetry $\tau$  of the Dynkin diagram,
corresponding to an exterior automorphism of the algebra
$\mathcal{G}$. In the finite case all the Dynkin diagrams of
simply-laced algebras show an automorphism of order $k = 2$,
except the case of D$_{4}$  where the order is 3. Let $\alpha_{i}$
be a simple root of $\mathcal{G}$. Using the automorphism  $\tau$
of order $k$, we obtain \be
\beta_{i}:=\alpha_{i}+\tau(\alpha_{i})+\cdots+\tau^{k-1}(\alpha_{i})
\label{eq:fold} \ee
 which form the simple root  system of
  a singular subalgebra $\mathcal{H}$ of $\mathcal{G}$.
 The generators of $\mathcal{G}$, corresponding to the simple roots $\alpha_{i}$, left
 unchanged by $\tau$, become generators of $\mathcal{H}$, while the
 other ones transform according to a relation analogous to
 eq.(\ref{eq:fold}). In the following we apply the folding method to
 the triple extended Lie algebras, obtaining  all the non simply laced triple extended algebras, as
in the finite case.
 The automorphism of the $3$-extended Dynkin diagram acts on the
 standard way upon the roots of the finite classical subalgebras and
 trivially on the extended roots $\alpha_{+1}, \alpha_{+2},
\alpha_{+3}$. This property has to hold if we want to preserve the
structure of the triple extension of the non-simply laced algebras
\footnote{Indeed, also the non simply-laced $G^{++}$ algebras can
be obtained with the same folding technique from the simply-laced
$\mathcal{G}^{++}$, while a different kind of folding applied to
the $\mathcal{G}^{+}$ algebras allows to get all the twisted
affine algebras.}.
 Let us enumerate all the cases.

\subsection{A$_{2N-1}^{+++} \longrightarrow$ C$_{N}^{+++}$}

\begin{center}
    \begin{picture}(180,20) \thicklines
        \put(0,0){\circle{14}}
        \put(84,0){\circle{14}}
        \put(168,0){\circle{14}}
        \put(84,-44){\circle{14}}
        \put(84,-84){\circle{14}}
        \put(84,-124){\circle{14}}
        \put(0,15){\makebox(0,0){$\alpha_{1}$}}
        \put(84,15){\makebox(0,0){$\alpha_{N}$}}
        \put(168,15){\makebox(0,0){$\alpha_{2N-1}$}}
        \put(68,-46){\makebox(0,0){$\alpha_{+1}$}}
        \put(68,-88){\makebox(0,0){$\alpha_{+2}$}}
        \put(68,-128){\makebox(0,0){$\alpha_{+3}$}}
        \put(5,-5){\line(2,-1){72}}
        \put(163,-5){\line(-2,-1){72}}
        \put(7,0){\line(1,0){4}}\put(15,0){\line(1,0){4}}
        \put(23,0){\line(1,0){4}}\put(31,0){\line(1,0){4}}
        \put(39,0){\line(1,0){4}}\put(47,0){\line(1,0){4}}
        \put(55,0){\line(1,0){4}}\put(63,0){\line(1,0){4}}
        \put(71,0){\line(1,0){4}}
        \put(91,0){\line(1,0){4}}\put(99,0){\line(1,0){4}}
        \put(107,0){\line(1,0){4}}\put(115,0){\line(1,0){4}}
        \put(123,0){\line(1,0){4}}\put(131,0){\line(1,0){4}}
        \put(139,0){\line(1,0){4}}\put(147,0){\line(1,0){4}}
        \put(155,0){\line(1,0){4}}
        \put(84,-51){\line(0,-1){25}}
        \put(84,-91){\line(0,-1){25}}
    \end{picture}
\end{center}

\vspace{4,5cm}

 The Cartan matrix of A$_{2N-1}^{+++}$ can be written, in block form,
 as
\begin{equation}
A=(a_{ij})=(\alpha_{i},\alpha_{j})= \left(\begin{array}{ccccccc}
 &  &  & -1 & 0 & 0\\
 & A_{A_{2N-1}} &  & \vdots & \vdots & \vdots\\
 &  &  & -1 & 0 & 0\\
-1 & \cdots & -1 & 2 & -1 & 0\\
0 &  \cdots & 0 & -1 & 2 & -1\\
0 &  \cdots & 0 & 0 & -1 & 2\\
\end{array}\right)
\end{equation}

The not trivial action of $\tau$ on the simple roots gives \bea
\beta_{1} & :=
&\alpha_{1}+\tau(\alpha_{1}) = \alpha_{1}+\alpha_{2N-1} \quad \\
\beta_{2} & := &
\alpha_{2}+\tau(\alpha_{2})=\alpha_{2}+\alpha_{2N-2}\quad\ldots
\\
\beta_{N-1}& := &
\alpha_{N-1}+\tau(\alpha_{N-1}) = \alpha_{N-1}+\alpha_{N+1} \quad \\
\beta_{N} & := & \alpha_{N}+\tau(\alpha_{N}) = 2\alpha_{N} \\
\beta_{+1} & := & \alpha_{+1}+\tau(\alpha_{+1}) =
2\alpha_{+1} \quad \\
\beta_{+2} & := & \alpha_{+2}+\tau(\alpha_{+2}) = 2\alpha_{+2}\quad \\
\beta_{+3} & := & \alpha_{+3}+\tau(\alpha_{+3}) = 2\alpha_{+3}
\eea
the length of the simple roots is \bea
 \beta^{2}_{1} & = & \ldots = \beta^{2}_{N-1}=4  \\
\beta^{2}_{+3} & = & \beta^{2}_{+2} =\beta^{2}_{+1 }=
\beta^{2}_{N} = 8\,. \eea
 The corresponding Cartan matrix is
\begin{equation}
B=(b_{ij})_{i,j}=2\frac{(\beta_{i},\beta_{j})}{(\beta_{i},\beta_{i})}=
\left(\begin{array}{ccccccc}
 &  &  & -2 & 0 & 0\\
 & A_{C_{N}} &  & \vdots & \vdots & \vdots\\
 &  &  & 0 & 0 & 0\\
-1 & \cdots & 0 & 2 & -1 & 0\\
0 &  \cdots & 0 & -1 & 2 & -1\\
0 &  \cdots & 0 & 0 & -1 & 2\\
\end{array}\right)
\end{equation}
\newline  So we get the $3$-extended Lie algebra C$_{N}^{+++}$ with
  Dynkin diagram
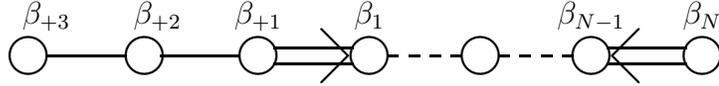
\begin{center}
    \begin{picture}(30,20) \thicklines
        \multiput(0,0)(42,0){5}{\circle{14}}
        \put(0,15){\makebox(0,0){$\beta_{+1}$}}
        \put(42,15){\makebox(0,0){$\beta_{1}$}}
        \put(126,15){\makebox(0,0){$\beta_{N-1}$}}
        \put(168,15){\makebox(0,0){$\beta_{N}$}}
        \put(6,-3){\line(1,0){30}}
        \put(6,3){\line(1,0){30}}
        \put(34,0){\line(-1,1){10}}\put(34,0){\line(-1,-1){10}}
        \put(49,0){\line(1,0){4}}\put(57,0){\line(1,0){4}}
        \put(65,0){\line(1,0){4}}\put(73,0){\line(1,0){4}}
        \put(91,0){\line(1,0){4}}\put(99,0){\line(1,0){4}}
        \put(107,0){\line(1,0){4}}\put(115,0){\line(1,0){4}}
        \put(132,-3){\line(1,0){30}}
        \put(132,3){\line(1,0){30}}
        \put(-7,0){\line(-1,0){30}}
        \put(-43,0){\circle{14}}
        \put(-38,15){\makebox(0,0){$\beta_{+2}$}}
        \put(-87,0){\circle{14}}
        \put(-50,0){\line(-1,0){30}}
        \put(-80,15){\makebox(0,0){$\beta_{+3}$}}
        \put(134,0){\line(1,-1){10}}\put(134,0){\line(1,1){10}}
    \end{picture}
\end{center}

\subsection{D$_{N}^{+++} \longrightarrow$ B$_{N-1}^{+++}$}

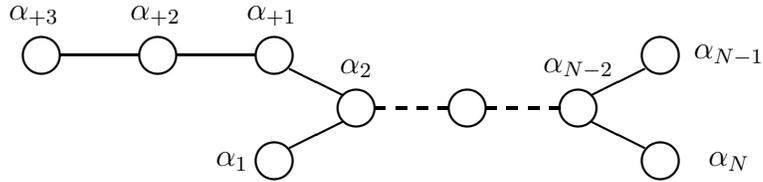
\bigskip
\begin{center}
   \begin{picture}(30,30) \thicklines
       \multiput(42,0)(42,0){3}{\circle{14}}
       \put(11,20){\circle{14}}
       \put(10,35){\makebox(0,0){$\alpha_{+1}$}}
       \put(11,-20){\circle{14}}
       \put(-5,-20){\makebox(0,0){$\alpha_{1}$}}
       \put(37,5){\line(-2,1){20}}\put(37,-5){\line(-2,-1){20}}
       \put(42,15){\makebox(0,0){$\alpha_{2}$}}
       \put(126,15){\makebox(0,0){$\alpha_{N-2}$}}
       \put(49,0){\line(1,0){4}}\put(57,0){\line(1,0){4}}
       \put(65,0){\line(1,0){4}}\put(73,0){\line(1,0){4}}
       \put(91,0){\line(1,0){4}}\put(99,0){\line(1,0){4}}
       \put(107,0){\line(1,0){4}}\put(115,0){\line(1,0){4}}
       \put(131,5){\line(2,1){20}}\put(131,-5){\line(2,-1){20}}
       \put(157,20){\circle{14}}
       \put(183,20){\makebox(0,0){$\alpha_{N-1}$}}
       \put(157,-20){\circle{14}}
       \put(183,-20){\makebox(0,0){$\alpha_{N}$}}
       \put(-34,35){\makebox(0,0){$\alpha_{+2}$}}
       \put(-33,20){\circle{14}}
       \put(4,20){\line(-1,0){30}}
       \put(-80,35){\makebox(0,0){$\alpha_{+3}$}}
       \put(-77,20){\circle{14}}
       \put(-40,20){\line(-1,0){30}}
   \end{picture}
\end{center}

\bigskip

The Cartan matrix can be written as
\begin{equation}
A=(a_{ij})_{i,j}=(\alpha_{i},\alpha_{j})=
\left(\begin{array}{ccccccc}
 &  &  &  & 0 & 0 & 0\\
 &  &  &  & -1 & \vdots & \vdots\\
 &  & A_{D_{N}} &  & \vdots & \vdots & \vdots\\
 &  &  &  & 0 & 0 & 0\\
0 & -1 & \cdots & 0 & 2 & -1 & 0\\
0 & \cdots & \cdots & 0 & -1 & 2 & -1\\
0 & \cdots & \cdots & 0 & 0 & -1 & 2\\
\end{array}\right)
\end{equation}
 The non trivial action of $\tau$ is
\begin{equation}
\tau(\alpha_{N-1})=\alpha_{N},\quad\tau(\alpha_{N})=\alpha_{N-1}
\end{equation}
 The new simple roots are
 \bea
\beta_{N-1} & := &
\alpha_{N-1}+\tau(\alpha_{N-1})=\alpha_{N-1}+\alpha_{N}\quad \\
\beta_{N-2} & := &
\alpha_{N-2}+\tau(\alpha_{N-2})=2\alpha_{N-2}\quad\ldots \\
\beta_{1} & := & \alpha_{1}+\tau(\alpha_{1})=2\alpha_{1} \quad \\
\beta_{+1}& := &
\alpha_{+1}+\tau(\alpha_{+1})=2\alpha_{+1} \quad \\
 \beta_{+2} & := & \alpha_{+2}+\tau(\alpha_{+2})=2\alpha_{+2} \quad \\
\beta_{+3} & := & \alpha_{+3}+\tau(\alpha_{+3})=2\alpha_{+3}
 \eea
 where
 \be
 \beta^{2}_{N-1} = 4\,, \;\;\;\;\;\; \beta^{2}_{+3} = \beta^{2}_{+2} = \ldots = \beta^{2}_{N-2} =
8\,. \ee
 The Cartan matrix and the corresponding Dynkin diagram of
B$_{N-1}^{+++}$  are
\begin{equation}
B=(b_{ij})_{i,j}=2\,\frac{(\beta_{i},\beta_{j})}{(\beta_{i},\beta_{i})}=
\left(\begin{array}{ccccccc}
 &  &  &  & 0 & 0 & 0\\
 &  &  &  & -1 & \vdots & \vdots\\
 &  & A_{B_{N-1}} &  & \vdots & \vdots & \vdots\\
 &  &  &  & 0 & 0 & 0\\
0 & -1 & \cdots & 0 & 2 & -1 & 0\\
0 & \cdots & \cdots & 0 & -1 & 2 & -1\\
0 & \cdots & \cdots & 0 & 0 & -1 & 2\\
\end{array}\right)
\end{equation}

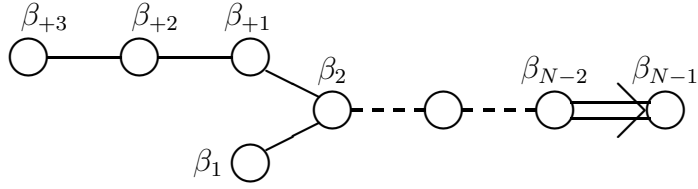
\begin{center}
    \begin{picture}(60,50) \thicklines
        \multiput(42,0)(42,0){4}{\circle{14}}
        \put(11,20){\circle{14}}
        \put(10,35){\makebox(0,0){$\beta_{+1}$}}
        \put(11,-20){\circle{14}}
        \put(-5,-20){\makebox(0,0){$\beta_{1}$}}
        \put(37,5){\line(-2,1){20}}\put(37,-5){\line(-2,-1){20}}
        \put(42,15){\makebox(0,0){$\beta_{2}$}}
        \put(126,15){\makebox(0,0){$\beta_{N-2}$}}
        \put(168,15){\makebox(0,0){$\beta_{N-1}$}}
        \put(49,0){\line(1,0){4}}\put(57,0){\line(1,0){4}}
        \put(65,0){\line(1,0){4}}\put(73,0){\line(1,0){4}}
        \put(91,0){\line(1,0){4}}\put(99,0){\line(1,0){4}}
        \put(107,0){\line(1,0){4}}\put(115,0){\line(1,0){4}}
        \put(132,-3){\line(1,0){30}}
        \put(132,3){\line(1,0){30}}
        \put(-24,20){\line(1,0){28}}
        \put(-31,20){\circle{14}}
        \put(-25,35){\makebox(0,0){$\beta_{+2}$}}
        \put(-38,20){\line(-1,0){28}}
        \put(-73,20){\circle{14}}
        \put(-67,35){\makebox(0,0){$\beta_{+3}$}}
        \put(160,0){\line(-1,1){10}}\put(160,0){\line(-1,-1){10}}
    \end{picture}
\end{center}

 \bigskip

\subsection{E$_{6}^{+++} \longrightarrow$ F$_{4}^{+++}$}

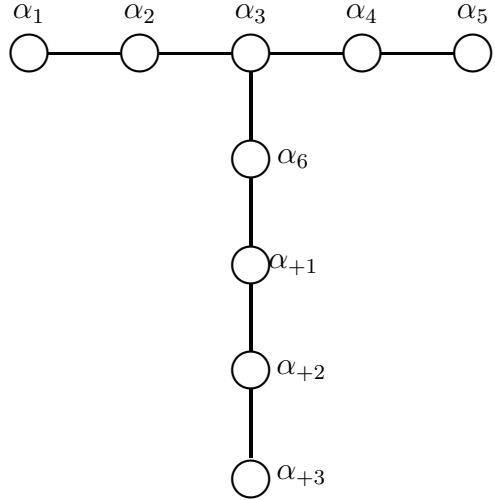
\begin{center}
    \begin{picture}(180,40) \thicklines
        \multiput(0,20)(42,0){5}{\circle{14}}
        \put(0,35){\makebox(0,0){$\alpha_{1}$}}
        \put(42,35){\makebox(0,0){$\alpha_{2}$}}
        \put(84,35){\makebox(0,0){$\alpha_{3}$}}
        \put(126,35){\makebox(0,0){$\alpha_{4}$}}
        \put(168,35){\makebox(0,0){$\alpha_{5}$}}
        \put(7,20){\line(1,0){28}}
        \put(49,20){\line(1,0){28}}
        \put(91,20){\line(1,0){28}}
        \put(133,20){\line(1,0){28}}
        \put(84,13){\line(0,-1){26}}
        \put(84,-20){\circle{14}}
        \put(100,-20){\makebox(0,0){$\alpha_{6}$}}
        \put(84,-27){\line(0,-1){26}}
        \put(84,-60){\circle{14}}
        \put(100,-60){\makebox(0,0){$\alpha_{+1}$}}
        \put(84,-67){\line(0,-1){26}}
        \put(84,-100){\circle{14}}
        \put(84,-107){\line(0,-1){26}}
        \put(84,-141){\circle{14}}
        \put(103,-100){\makebox(0,0){$\alpha_{+2}$}}
        \put(103,-140){\makebox(0,0){$\alpha_{+3}$}}
    \end{picture}
\end{center}

\vspace{5cm}

 The Cartan matrix of E$_{6}^{+++}$ is

\begin{equation}
A=(a_{ij})_{i,j}=(\alpha_{i},\alpha_{j})=
\left(\begin{array}{cccccc}
 &  &  & 0 & 0 & 0\\
 & A_{E_{6}} &  & \vdots & \vdots & \vdots\\
 &  &  & -1 & 0 & 0\\
0 & \cdots & -1 & 2 & -1 & 0\\
0 & \cdots & 0 & -1 & 2 & -1\\
0 & \cdots & 0 & 0 & -1 & 2\\
\end{array}\right)
\end{equation}
The simple roots are given by \bea \beta_{1} & := &
\alpha_{1}+\tau(\alpha_{1})=\alpha_{1}+\alpha_{5}
\quad \\
\beta_{2} & := & \alpha_{2}+\tau(\alpha_{2})=\alpha_{2}+\alpha_{4}
\quad \\
\beta_{3} & := & \alpha_{3}+\tau(\alpha_{3})=2\alpha_{3} \quad \\
\beta_{4} & := & \alpha_{6}+\tau(\alpha_{6})=2\alpha_{6}\quad \\
\beta_{+1} & := & 2\alpha_{+1},\quad\beta_{+2}:=2\alpha_{+2}\quad \\
\beta_{+3} & := & 2\alpha_{+3} \eea and \be
 \beta^{2}_{1} = \beta^{2}_{2}  = 4\,, \;\;\;\;\;\;
\beta^{2}_{+3} = \beta^{2}_{+2} = \ldots = \beta^{2}_{3} = 8\,.
\ee

One gets the $3$-extended algebra F$_{4}^{+++}$, with Cartan
matrix and Dynkin diagram

\begin{equation}
B=(b_{ij})_{i,j}=2\,\frac{(\beta_{i},\beta_{j})}{(\beta_{i},\beta_{i})}=
\left(\begin{array}{cccccc}
 &  &  & 0 & 0 & 0\\
 & A_{F_{4}} &  & \vdots & \vdots & \vdots\\
 &  &  & -1 & 0 & 0\\
0 & \cdots & -1 & 2 & -1 & 0\\
0 & \cdots & 0 & -1 & 2 & -1\\
0 & \cdots & 0 & 0 & -1 & 2\\
\end{array}\right)
\end{equation}

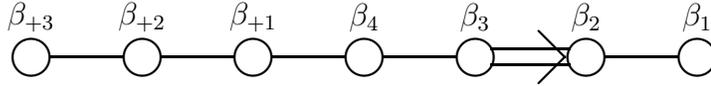
\begin{center}
    \begin{picture}(30,20) \thicklines
        \multiput(-84,0)(42,0){7}{\circle{14}}
        \put(0,15){\makebox(0,0){$\beta_{+1}$}}
        \put(42,15){\makebox(0,0){$\beta_{4}$}}
        \put(84,15){\makebox(0,0){$\beta_{3}$}}
        \put(126,15){\makebox(0,0){$\beta_{2}$}}
        \put(168,15){\makebox(0,0){$\beta_{1}$}}
        \put(7,0){\line(1,0){28}}
        \put(49,0){\line(1,0){28}}
        \put(90,-3){\line(1,0){30}}
        \put(90,3){\line(1,0){30}}
        \put(133,0){\line(1,0){28}}
        \put(118,0){\line(-1,1){10}}
        \put(118,0){\line(-1,-1){10}}
        \put(-42,15){\makebox(0,0){$\beta_{+2}$}}
        \put(-84,15){\makebox(0,0){$\beta_{+3}$}}
        \put(-7,0){\line(-1,0){28}}
        \put(-49,0){\line(-1,0){28}}
    \end{picture}
\end{center}

\subsection{$D_{4}^{+++} \longrightarrow G_{2}^{+++}$}

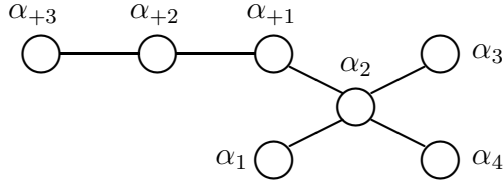
\bigskip
\begin{center}
   \begin{picture}(0,30) \thicklines
       \multiput(42,0)(42,0){1}{\circle{14}}
       \put(11,20){\circle{14}}
       \put(10,35){\makebox(0,0){$\alpha_{+1}$}}
       \put(11,-20){\circle{14}}
       \put(-5,-20){\makebox(0,0){$\alpha_{1}$}}
       \put(37,5){\line(-2,1){20}}\put(37,-5){\line(-2,-1){20}}
       \put(42,15){\makebox(0,0){$\alpha_{2}$}}
       \put(48,5){\line(2,1){20}}\put(48,-5){\line(2,-1){20}}
       \put(74,20){\circle{14}}
       \put(92,20){\makebox(0,0){$\alpha_{3}$}}
       \put(74,-20){\circle{14}}
       \put(92,-20){\makebox(0,0){$\alpha_{4}$}}
       \put(-34,35){\makebox(0,0){$\alpha_{+2}$}}
       \put(-33,20){\circle{14}}
       \put(4,20){\line(-1,0){30}}
       \put(-80,35){\makebox(0,0){$\alpha_{+3}$}}
       \put(-77,20){\circle{14}}
       \put(-40,20){\line(-1,0){30}}
   \end{picture}
\end{center}

\bigskip
\bigskip

 The Cartan matrix of  D$_{4}^{+++}$ is
\begin{equation}
A=(a_{ij})_{i,j}=(\alpha_{i},\alpha_{j})=
\left(\begin{array}{ccccccc}
2 & -1 & 0 & 0 & 0 & 0 & 0\\
-1 & 2 & -1 & -1 & -1 & 0 & 0\\
0 & -1 & 2 & 0 & 0 & 0 & 0\\
0 & -1 & 0 & 2 & 0 & 0 & 0\\
0 & -1 & 0 & 0 & 2 & -1 & 0\\
0 & 0 & 0 & 0 & -1 & 2 & -1\\
0 & 0 & 0 & 0 & 0 & -1 & 2\\
\end{array}\right)
\end{equation}
The action of $\tau$ on the simple roots gives \bea \beta_{1} & :=
& \alpha_{1}+\tau(\alpha_{1})+\tau^{2}(\alpha_{1}) =
\alpha_{1}+\alpha_{3}+\alpha_{4}\quad \\
\beta_{2} & := & 3\alpha_{2} \quad \\
\beta_{+1} & := & 3\,\alpha_{+1}\,,\quad \beta_{+2}:= 3\,\alpha_{+2} \quad \\
\beta_{+3}& := & 3\,\alpha_{+3}
 \eea
 and
 \be
 \beta^{2}_{1}=6\,,  \;\;\;\;\;
\beta^{2}_{+3}=\ldots=\beta^{2}_{2}=18\,. \ee
One gets the
extended algebra G$_{2}^{+++}$  with Cartan  matrix and  Dynkin
diagram:

\begin{equation}
B=(b_{ij})_{i,j}=2\,\frac{(\beta_{i},\beta_{j})}{(\beta_{i},\beta_{i})}=
\left(\begin{array}{ccccc}
2 & -3 & 0 & 0 & 0\\
-1 & 2 & -1 & 0 & 0\\
0 & -1 & 2 & -1 & 0\\
0 & 0 & -1 & 2 & -1\\
0 & 0 & 0 & -1 & 2
\end{array}\right)
\end{equation}

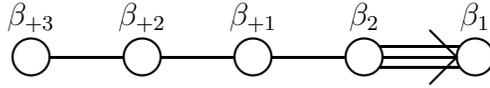
\begin{center}
    \begin{picture}(30,20) \thicklines
        \multiput(-84,0)(42,0){5}{\circle{14}}
        \put(0,15){\makebox(0,0){$\beta_{+1}$}}
        \put(42,15){\makebox(0,0){$\beta_{2}$}}
        \put(84,15){\makebox(0,0){$\beta_{1}$}}
        \put(-42,15){\makebox(0,0){$\beta_{+2}$}}
        \put(-84,15){\makebox(0,0){$\beta_{+3}$}}
        \put(7,0){\line(1,0){28}}
        \put(48,-4){\line(1,0){30}}
        \put(49,0){\line(1,0){28}}
        \put(48,4){\line(1,0){30}}
        \put(77,0){\line(-1,1){10}}
        \put(77,0){\line(-1,-1){10}}
        \put(-7,0){\line(-1,0){28}}
        \put(-49,0){\line(-1,0){28}}
    \end{picture}
\end{center}

\bigskip

Let us remark that, in general, by applying the folding procedure
to an algebra $\mathcal{G}$, defined by the Cartan matrix $A$ and
Dynkin diagram $S(A)$, we obtain the Cartan matrix $B$ and the
corresponding Dynkin diagram $S(B)$ of another algebra
$\mathcal{H}$.
    However we have to check that the new generators, defined in
    function of the generators of of $\mathcal{G}$,
satisfy all the defining relations of the algebra  $\mathcal{H}$.
Let us see how the new generators are obtained.
 Let $\alpha_{i}$ be a simple root of of $\mathcal{G}$ and $h_{i}, e_{i}, f_{i}$
the associated generators. Let us denote by $\beta_{i}$,
respectively $h'_{i}, e'_{i}, f'_{i}$,
 the  roots and the associated generators transformed under the action of $\tau$, which we
 identify, respectively, as the
 simple roots and the  associated generators of $\mathcal{H}$.
 If the action of $\tau$ is trivial, that is $\beta_{i} = k \alpha_{i}$ (where $k$ is the order of the automorphism $\tau$, $\tau^{k}=1$)
 then the generators are not transformed
$h'_{i}=h_{i}$, $e'_{i}=e_{i}$ and $f'_{i}=f_{i}$.  If the action
of $\tau$ is not trivial, that is
 $\beta_{i}=\alpha_{i} + \tau(\alpha_{i}) + \ldots +
\tau^{k-1}(\alpha_{i})$, we obtain $h'_{i}=h_{i} +
h_{\tau(\alpha_{i})} + \ldots + h_{\tau^{k-1}(\alpha_{i})}$,
$e'_{i}=e_{i} + e_{\tau(\alpha_{i})} + \ldots +
e_{\tau^{k-1}(\alpha_{i})}$ and $f'_{i}=f_{i} +
f_{\tau(\alpha_{i})} + \ldots + f_{\tau^{k-1}(\alpha_{i})}$.  We
have to verify that the generators $h'_{i}, e'_{i}, f'_{i}$
satisfy the defining relations
$$
[e'_{i},f'_{j}]=\delta_{ij}\,h'_{j},
$$
$$
[h'_{i},e'_{j}]=b_{ij}\,e'_{j},\quad[h'_{i},f'_{j}]=-b_{ij}\,f'_{j},
$$
$$
[h'_{i},h'_{j}]=0,
$$
$$
(\mbox{ad}\: e'_{i})^{1-b_{ij}}\:e'_{j}=0,\quad(\mbox{ad}\:
f'_{i})^{1-b_{ij}}\:f'_{j}=0,\; (i\neq j) .
$$
We do no report here the explicit calculations, but everything
works nicely. Finally it should be remarked that the folding
procedure for indefinite Kac-Moody algebra, when applicable,
always gives rise to indefinite Kac-Moody algebra,  as it happens
for the finite, affine, hyperbolic Kac-Moody algebras. On the
contrary other reduction procedures, as the orbifolding, do not
preserve the kind of algebras. Indeed as remarked in \cite{Helf},
the orbifolding of $E_{10}$, to which the folding procedure cannot
be applied, gives rise to non Kac-Moody algebras.

\section{Non-standard extensions of Lie algebras} \label{nonstandard}

In this section we present a non-standard construction of extended
Lie algebras; as stated in Sec. 1, the idea of the non-standard
extension is to add to the simple root system  \{$\alpha_{i}$\} of
a simple Lie algebra $\mathcal{G}$ new roots, which are formed by
those fundamental weights of the algebra that are linear
combinations with integer coefficients  of  $\alpha_{i}$, plus a
suitable combinations of vectors belonging to the Lorentzian
lattice II$^{1,1}_{k_{\pm}}$ and/or  II$^{1,1}_{l_{\pm}}$. The new
roots have to satisfy the requirements that their squared norms
are equal to 2 and that are suitably  linked  with the previous
ones. Let us remark that the roots of the non-standard extension
do not generally span the whole lattice ${\bf \Lambda} =
\Lambda_{\mathcal{G}}\, \oplus  \mbox{II}^{1,1}_{k_{\pm}} \oplus
\mbox{II}^{1,1}_{l_{\pm}}$ and that,  moreover, the structure of
the added simple root is, by no way, unique.  Of course one  can
add more than two $2$-dim. Lorentzian lattices, but these
extensions will not be considered in the present paper, where we
add at most three new roots.  Also we shall not discuss the case
where the squared norm of the added roots is not equal to 2. So,
given a simple Lie algebra $\mathcal{G}$, we add to the root
lattice $\Lambda_{\mathcal{G}}$ a new simple root $\alpha_{r+1}
\equiv \alpha_{+1}$, which is formed by the opposite of a
fundamental weight $- \Lambda_{i}$  and by a suitable linear
combination, with integer coefficients, of the vectors $k_{\pm}$,
in order to have $\alpha_{+1}^{2}=2$, as $\Lambda_{i}^{2}$ is not
necessarily 2. Let's remember that the fundamental weights have
the property
$2\frac{(\Lambda_{i},\alpha_{j})}{(\alpha_{j},\alpha_{j})} =
\delta_{i,j}$; they span the weight lattice $P =
\bigoplus_{i=1}^{r} \mathbb{Z} \Lambda_{i}$, which is dual to the
coroot lattice $\Lambda_{\mathcal{G}}^{\vee} = \bigoplus_{i=1}^{r}
\mathbb{Z} \alpha_{i}^{\vee}$ where $\alpha_{i}^{\vee}=2
\frac{\alpha_{i}}{(\alpha_{i},\alpha_{i})}$ are the coroots. So
while it is always true that
$P=(\Lambda_{\mathcal{G}}^{\vee})^{\ast}$, in general we have
$\Lambda_{\mathcal{G}} \subseteq P$. This means that for each
$\mathcal{G}$, only some $\Lambda_{i}$ belong to the root lattice;
so, in defining $\alpha_{+1}$ we choose the $\Lambda_{i} \in
\Lambda_{\mathcal{G}}$. Since $\Lambda_{i}$ is only linked with
the simple root $\alpha_{i}$, we have
$(\alpha_{+1},\alpha_{j})=-\delta_{+1,\,i}$. So the first
extension is made by adding the root
  $\alpha_{r+1} \equiv \alpha_{+1}  \equiv \ := - \Lambda_{i} + k_{+}  - a k_{-}$,  where $ a \in
  {\mathbf{Z}_{+}}$ is fixed by the condition   $ \alpha_{+1}^{2} =  \Lambda_{i} ^{2} - 2 a = 2.$
  At this point, we add
the simple root $\alpha_{r+2} \equiv \alpha_{+2} := -\theta + b
k_{-} - l_{-}$, where $\theta$ is the highest root of
$\mathcal{G}$ and $b  \in \mathbf{Z}$ is a coefficient chosen in
order to have $(\alpha_{+2},\alpha_{+1}) =  ( \Lambda_{i}, \,
\theta) - b = 0$. In this way, we have $\alpha_{+2}^{2}=2$ and
$\alpha_{+2}$ behaves like an affine root (that is, it is linked
with the simple roots of $\mathcal{G}$ in a way completely
analogous as the affine root of the algebras
$\mathcal{\widehat{G}}$). At the end, we add the third simple root
$\alpha_{r+3} \equiv \alpha_{+3} := l_{+} + l_{-}$ with the
property that $(\alpha_{+3},\alpha_{+2}) = -1$ and
$(\alpha_{+3},\alpha_{i}) = 0$ for $i = 1,\ldots, r, r+1$. As
$(\Lambda_{i}, \, \Lambda_{j}) \in \mathbf{Z}_{>}$ for the Lie
algebra  $\mathcal{G}$ below considered, this procedure is
completely general and the extended algebra contains as subalgebra
the affine extension of $\mathcal{G}$ (so sometimes we shall call
this a {\textsl{non standard affine extension}). Clearly the
lightlike vector $l_{-}$ can be hanged up to any other simple
root, producing an other indefinite Kac-Moody. We shall comment on
this point in Sec.\ref{subalgebras}.
This construction leads to indefinite Kac-Moody algebras, 1-, 2-
and 3-extended, whose (symmetric) Cartan matrix $2
\frac{(\alpha_{i},\alpha_{j})}{(\alpha_{i},\alpha_{i})}$ (for $i,
j = 1, \ldots, r+3$) has Lorentzian signature $(+ \ldots + -)$
with $r+2$ plus signs and 1 minus sign. The root lattice of the
3-extended algebra is properly contained in
$\Lambda_{\mathcal{G}}\, \oplus$ II$^{1,1}_{k_{\pm}} \oplus$
II$^{1,1}_{l_{\pm}}$. For these algebras, one can make similar
discussions as those in Sec. \ref{triple} on eventual further
extensions. Now we want to discuss another possible extension,
which cannot be performed for any fundamental weight $\Lambda_{i}$
belonging to the root lattice of ${\mathcal{G}}$. The first
extension is performed as before, but as second extension we add
the root  $ \alpha_{+2} :=  - \Lambda_{j} + k_{+}  - b k_{-} -
l_{-}$ ($ i \neq j$),  where $ b \in {\mathbf{Z}_{+}}$ is such
that  $ \alpha_{+2}^{2} =  2$ and $ (\alpha_{+1},  \alpha_{+2}) =
(\Lambda_{i}, \Lambda_{j}) - a - b = 0$. Below we shall show that,
for   ${\mathcal{G}} \neq E_{6}$, for any $i$ ($ \Lambda_{i} \neq
\theta$), at least one $j$ exists which satisfies the above
condition.
In the following, we discuss only some examples of the general
construction, we called \textsl{affine extension}; in particular
we concentrate on the simply laced algebras, but it is possible to
consider also the other cases paying attention at the choice of
the fundamental weight.

Looking at the fundamental weights of simply laced-Lie algebras,
see  \cite{FSS}, one realizes that  the fundamental weights  which
can be written as \be \Lambda_{i} = \sum_{n} \, c_{n} \,
\alpha_{n} \;\;\;\;\;\;\;\;\;\;  c_{n}  \in {\bf Z} \ee are
\begin{enumerate}
\item for $D_{N} = so(2N)$ ($N \geq 4$), the weights $\Lambda_{i}$
with $i$ even number ($N-2 \geq i \geq 2$)
 \item for $E_{6}$, only the weights $\Lambda_{i}$ ($i = 3,6$)
 \item for  $E_{7}$, only the weights $\Lambda_{i}$ ($i = 1,2,3,5$)
 \item for $E_{8}$,  all the weights
$\Lambda_{i}$, which is just a consequence of the $E_{8}$-lattice
being a self-dual one.
 \end{enumerate}
  In the following we discuss some of the
possible non-standard extensions, with the aim to illustrate the
procedure in a few examples which may be relevant for their
subalgebras content.  Let us emphasize that the discussed
extensions as well their subalgebras content are not at all
exhaustive,  being the choice of the extended simple roots not
unique,  in general.

\subsection{$D_{N} = so(2N)$}

In order to illustrate the general procedure, we discuss in some
detail the case of $D_{6}=so(12)$, which is the first algebra of
the even orthogonal series which admits a non-standard extension.
We add the simple root \be \alpha_{+1} := - \Lambda_{4} + k_{+} -
k_{-}, \;\;\;\;\;\;\;\;\;\;\;\; \alpha_{+1}^{2} = 2, \ee where
$\Lambda_{4}$ is the   fundamental weight \footnote{the
$\varepsilon_{i}$ are unit ortho-normal vectors in
$\mathbb{R}^{6}$.}

\be \Lambda_{4} = \eps_{1} + \eps_{2} + \eps_{3} + \eps_{4},
\;\;\;\;\;\;\;\;\;\;\;\; \Lambda_{3}^{2} = 4, \ee Clearly we have
\be (\alpha_{+1}, \alpha_{i}) = - \delta_{4,\,i}. \ee We add now
the root \be \alpha_{+2}  := -h.r.- 2 k_{-}  - l_{-} = -(\eps_{1}
+ \eps_{2})   - 2 k_{-}  -  l_{-}, \ee
\be (\alpha_{+2}, \alpha_{j}) = - \delta_{2,\,j}, \ee and \be
\alpha_{+3}  :=  l_{+} + l_{-}, \ee \be (\alpha_{+3}, \alpha_{k})
= - \delta_{+2,k}, \ee
with Dynkin diagram:

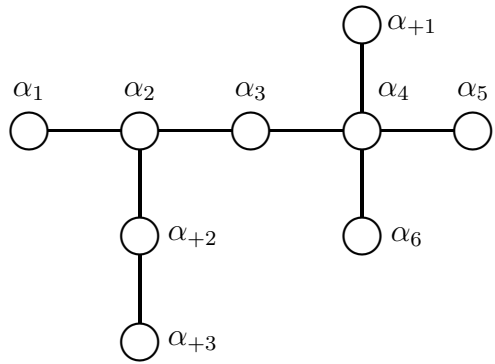
\vspace{1cm}
\begin{center}
    \begin{picture}(180,40) \thicklines
        \multiput(0,20)(42,0){5}{\circle{14}}
        \put(0,35){\makebox(0,0){$\alpha_{1}$}}
        \put(42,35){\makebox(0,0){$\alpha_{2}$}}
        \put(84,35){\makebox(0,0){$\alpha_{3}$}}
        \put(138,35){\makebox(0,0){$\alpha_{4}$}}
        \put(168,35){\makebox(0,0){$\alpha_{5}$}}
        \put(7,20){\line(1,0){28}}
        \put(49,20){\line(1,0){28}}
        \put(91,20){\line(1,0){28}}
        \put(133,20){\line(1,0){28}}
        \put(126,13){\line(0,-1){26}}
        \put(42,13){\line(0,-1){26}}
        \put(42,-20){\circle{14}}
        \put(62,-20){\makebox(0,0){$\alpha_{+2}$}}
        \put(42,-27){\line(0,-1){26}}
        \put(42,-60){\circle{14}}
        \put(62,-60){\makebox(0,0){$\alpha_{+3}$}}
        \put(126,-20){\circle{14}}
        \put(126,27){\line(0,1){26}}
        \put(126,60){\circle{14}}
        \put(145,60){\makebox(0,0){$\alpha_{+1}$}}
        \put(143,-20){\makebox(0,0){$\alpha_{6}$}}
    \end{picture}
\end{center}
\vspace{2cm}

Let's observe that the first extension of $D_{6}$ is the same
algebra as $D_{4}^{+++}$, so folding $\alpha_{+1}$, $\alpha_{5}$
and $\alpha_{6}$ we re-obtain $G_{2}^{+++}$. Clearly the choice of
the extended simple roots is not unique.  One can easily see that:
\begin{itemize}
\item  a non-standard extension of $so(4N)$ admits as subalgebra
the affine extension of  $so(4N)$ and $so(4(N-1))$. Indeed one
adds to the roots of $so(4N)$ the root of the affine extension \be
\alpha_{+1}  := - \Lambda_{2} +  k_{+} \ee and the new
non-standard root \be \alpha_{+2}  :=  - \Lambda_{4} +  l_{+} -
l_{-} \ee Taking away the roots $\alpha_{+1}, \alpha_{j}$ ($j =
1,2$) one gets the algebra $so(4(N-1))^{(1)}$. Let us remark that
if we add the root
    \be
\alpha_{+2}  :=  - \Lambda_{6} +  l_{+} -  2 l_{-} \ee
    and then we take away the roots
$\alpha_{+1}, \alpha_{j}$ ($j = 1,\ldots,4$) one gets the algebra
$so(4(N-2))^{(1)}$. \item the  non standard extension of $so(24)$
is the smallest extension of the orthogonal series  which contains
as subalgebra $E_{11}$. Indeed adding to the roots of $so(24)$
the non standard root \be \alpha_{+1}  :=  - \Lambda_{8} +  k_{+}
- 3 k_{-} \ee and deleting $\alpha_{11}, \alpha_{12}$  one gets
$E_{11}$.
\end{itemize}
  Let us call $\widehat{\Lambda}_{2n} = -  \Lambda_{2n}  + k_{+}  - (n-1) k_{-}$, where $n \in {\mathbb{Z}_{+}}$ and $\Lambda_{2n} = \sum_{i=1}^{2n}  \, \vr_{i}$ is a fundamental weight.
  Clearly we have
  \be
 \widehat{\Lambda}_{2n}^{2} = 2  \;\; \;\; \;\; \;\;  (\widehat{\Lambda}_{2n}, \widehat{\Lambda}_{2n+2}) = 0
 \ee

\subsection{$E_{6}$}

Let us add to the simple root system of $E_{6}$ the root \be
\alpha_{+1}  := - \Lambda_{3} + k_{+} - 2 k_{-},
\;\;\;\;\;\;\;\;\;\;\;\; \alpha_{+1}^{2} = 2, \ee where
$\Lambda_{3}$ is the   fundamental weight \be \Lambda_{3} =
\eps_{3} + \eps_{4} + \eps_{5} + \eps_{8} - \eps_{7} - \eps_{6},
\;\;\;\;\;\;\;\;\;\;\;\; \Lambda_{3}^{2} = 6. \ee Clearly we have
\be (\alpha_{+1}, \alpha_{i}) = - \delta_{3,\,i}. \ee We add now
the root \be \alpha_{+2}  := -h.r. - 3 k_{-}  -  l_{-} =
-\frac{1}{2}(\eps_{1} + \eps_{2} + \eps_{3} + \eps_{4} + \eps_{5}
- \eps_{6} - \eps_{7} + \eps_{8}) - 3 k_{-}  -  l_{-}, \ee \be
(\alpha_{+2}, \alpha_{j}) = - \delta_{6,\,j}, \ee and \be
\alpha_{+3}  :=   l_{+} + l_{-},  \;\;\;\;\;\;\;\, (\alpha_{+3},
\alpha_{k}) = - \delta_{+2,\,k}. \ee Alternatively we can add the
roots \bea \alpha_{+2}  &  := &  k_{-} + k_{+} - l_{-},
\;\;\;\;\;\;\;\, (\alpha_{+2}, \alpha_{j}) = - \delta_{+1,\,j},
 \\
\alpha_{+3}  &  := &  l_{+} + l_{-},  \;\;\;\;\;\;\;\,
(\alpha_{+3}, \alpha_{k}) = - \delta_{+2,\,k}, \eea
and we obtain the same Dynkin diagram:

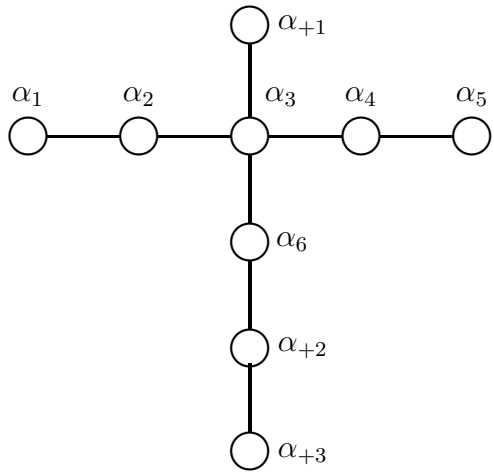
\vspace{1cm}
\begin{center}
    \begin{picture}(180,40) \thicklines
        \multiput(0,20)(42,0){5}{\circle{14}}
        \put(0,35){\makebox(0,0){$\alpha_{1}$}}
        \put(42,35){\makebox(0,0){$\alpha_{2}$}}
        \put(84,35){\makebox(0,0){\quad\quad $\alpha_{3}$}}
        \put(84,27){\line(0,1){28}}
        \put(84,62){\circle{14}}
        \put(104,62){\makebox(0,0){$\alpha_{+1}$}}
        \put(126,35){\makebox(0,0){$\alpha_{4}$}}
        \put(168,35){\makebox(0,0){$\alpha_{5}$}}
        \put(7,20){\line(1,0){28}}
        \put(49,20){\line(1,0){28}}
        \put(91,20){\line(1,0){28}}
        \put(133,20){\line(1,0){28}}
        \put(84,13){\line(0,-1){26}}
        \put(84,-20){\circle{14}}
        \put(100,-20){\makebox(0,0){$\alpha_{6}$}}
        \put(84,-27){\line(0,-1){26}}
        \put(84,-60){\circle{14}}
        \put(104,-60){\makebox(0,0){$\alpha_{+2}$}}
        \put(84,-66){\line(0,-1){26}}
        \put(84,-99){\circle{14}}
        \put(104,-100){\makebox(0,0){$\alpha_{+3}$}}
    \end{picture}
\end{center}
\vspace{3cm}
\bigskip
\bigskip
\bigskip
where in the second construction the roles of $\alpha_{+1}$ and
$\alpha_{6}$ are exchanged. Let's observe that many non-standard
(simply-laced) Dynkin diagrams can be folded to obtain other (non
simply-laced) Dynkin diagrams. For example, in the case of
$E_{6}$, we can identify the roots $\alpha_{1}$ and $\alpha_{2}$
with $\alpha_{5}$ and $\alpha_{4}$ respectively (so the new simple
roots are $\beta_{1}=\alpha_{1} + \alpha_{5}$,
$\beta_{2}=\alpha_{2} + \alpha_{4}$ and $\beta_{i} = 2\,
\alpha_{i}$ for $i = 3, +1, +2, +3 $), obtaining the following
folded Dynkin diagram:

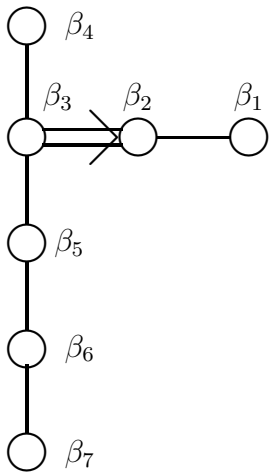
\vspace{1cm}
\begin{center}
    \begin{picture}(180,40) \thicklines
        \multiput(84,20)(42,0){3}{\circle{14}}
        \put(84,35){\makebox(0,0){\quad\quad $\beta_{3}$}}
        \put(84,27){\line(0,1){28}}
        \put(84,62){\circle{14}}
        \put(104,62){\makebox(0,0){$\beta_{4}$}}
        \put(126,35){\makebox(0,0){$\beta_{2}$}}
        \put(168,35){\makebox(0,0){$\beta_{1}$}}
        \put(90,17){\line(1,0){30}}
        \put(90,23){\line(1,0){30}}
        \put(118,20){\line(-1,-1){10}}
        \put(118,20){\line(-1,1){10}}
        \put(133,20){\line(1,0){28}}
        \put(84,13){\line(0,-1){26}}
        \put(84,-20){\circle{14}}
        \put(100,-20){\makebox(0,0){$\beta_{5}$}}
        \put(84,-27){\line(0,-1){26}}
        \put(84,-60){\circle{14}}
        \put(104,-60){\makebox(0,0){$\beta_{6}$}}
        \put(84,-66){\line(0,-1){26}}
        \put(84,-99){\circle{14}}
        \put(104,-100){\makebox(0,0){$\beta_{7}$}}
    \end{picture}
\end{center}
\bigskip
\vspace{3cm}

Actually, if we consider only the first extension of $E_{6}$, then
we can identify also $\alpha_{+1}$ with $\alpha_{6}$ and we
obtain:

\bigskip
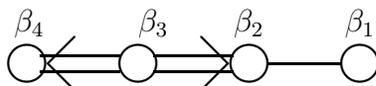
\begin{center}
    \begin{picture}(180,40) \thicklines
        \multiput(42,20)(42,0){4}{\circle{14}}
        \put(84,35){\makebox(0,0){\quad $\beta_{3}$}}
        \put(126,35){\makebox(0,0){$\beta_{2}$}}
        \put(168,35){\makebox(0,0){$\beta_{1}$}}
        \put(90,17){\line(1,0){30}}
        \put(90,23){\line(1,0){30}}
        \put(118,20){\line(-1,-1){10}}
        \put(118,20){\line(-1,1){10}}
        \put(133,20){\line(1,0){28}}
        \put(77,17){\line(-1,0){30}}
        \put(77,23){\line(-1,0){30}}
        \put(50,20){\line(1,1){10}}
        \put(50,20){\line(1,-1){10}}
        \put(43,35){\makebox(0,0){$\beta_{4}$}}
    \end{picture}
\end{center}

\subsection{$E_{7}$}

In this case we could use the fundamental weights $\Lambda_{1}$,
$\Lambda_{2}$, $\Lambda_{3}$ and $\Lambda_{5}$. To illustrate the
procedure, let's consider the weight $\Lambda_{5}$
($(\Lambda_{5},\Lambda_{5})=4$) and add the simple roots:

\begin{equation}
\alpha_{+1}  := - \Lambda_{5} + k_{+} - k_{-},
\;\;\;\;\;\;\;\;\;\;\;\; \alpha_{+1}^{2} = 2,  \quad\quad
(\alpha_{+1}, \alpha_{i}) = - \delta_{5,\,i},
\end{equation}
with
 \be
\Lambda_{5} =  \eps_{5} + \eps_{6} - \eps_{7} + \eps_{8},
\;\;\;\;\;\;\;\;\;\;\;\; \Lambda_{5}^{2} = 4. \ee
  Then
\begin{equation}
\alpha_{+2}  := - h.r. - 2 k_{-} - l_{-}\,  =  \eps_{7} -
\eps_{8} -2 k_{-} - l_{-}\, \quad\quad \alpha_{+3} = l_{+} +
l_{-}\, ,
\end{equation}
with $(\alpha_{+2},\alpha_{k})= - \delta_{k,\,1}$ and
$(\alpha_{+3},\alpha_{j})= - \delta_{j,\,+2}$. In this way we
obtain the Dynkin diagram:

\bigskip
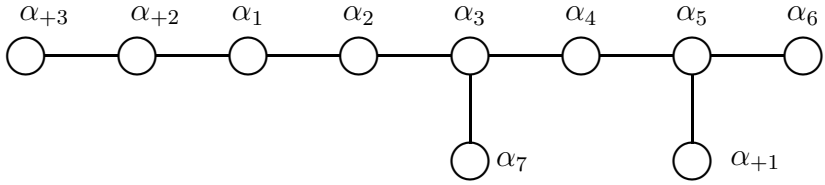
\begin{center}
    \begin{picture}(220,40) \thicklines
        \multiput(-84,20)(42,0){8}{\circle{14}}
        \put(0,35){\makebox(0,0){$\alpha_{1}$}}
        \put(42,35){\makebox(0,0){$\alpha_{2}$}}
        \put(84,35){\makebox(0,0){$\alpha_{3}$}}
        \put(126,35){\makebox(0,0){$\alpha_{4}$}}
        \put(168,35){\makebox(0,0){$\alpha_{5}$}}
        \put(210,35){\makebox(0,0){$\alpha_{6}$}}
        \put(-35,35){\makebox(0,0){$\alpha_{+2}$}}
        \put(-77,35){\makebox(0,0){$\alpha_{+3}$}}
        \put(-7,20){\line(-1,0){28}}
        \put(-49,20){\line(-1,0){28}}
        \put(7,20){\line(1,0){28}}
        \put(49,20){\line(1,0){28}}
        \put(91,20){\line(1,0){28}}
        \put(133,20){\line(1,0){28}}
        \put(175,20){\line(1,0){28}}
        \put(84,13){\line(0,-1){26}}
        \put(84,-20){\circle{14}}
        \put(168,13){\line(0,-1){26}}
        \put(168,-20){\circle{14}}
        \put(192,-20){\makebox(0,0){$\alpha_{+1}$}}
        \put(100,-20){\makebox(0,0){$\alpha_{7}$}}
    \end{picture}
\end{center}
\vspace{1cm}

  Let us call $\widehat{\Lambda}_{i}  := -  \Lambda_{i}  + k_+  - a k_-$, where $a \in {\mathbb{Z}_{+}}$ and $\Lambda_{i}$ , $ i = 1,2,3,5$, is a fundamental weight. We have
  \be
 \widehat{\Lambda}_{i}^{2} = 2  \;\; \;\; \;\; \;\;  (\widehat{\Lambda}_{3}, \widehat{\Lambda}_{5}) = 0
 \ee

\subsection{$E_{8}$}

$E_{8}$ root lattice is self-dual, so it coincides with the weight
lattice. The non-standard extension can be made adding to the
simple root system a root  equal  to the opposite of any weight
$\Lambda_{i}$ ($i=1, \ldots, 8$) plus some combination of $k_{+}$
and $k_{-}$. In this way we have 8 different extensions of
$E_{8}$, with the nodes $+2$, $+3$ always in the same position
(that is $(\alpha_{+2},\alpha_{7})= -1$ and
$(\alpha_{+3},\alpha_{+2})= -1$), while the node $+1$ moves from
the node 1 to the node 8, when $i$ runs from 1 to 8 respectively.
Actually, this situation is general for the non-standard
extensions.

Let us emphasize again that the choice of the extended simple
roots is not unique at all. Motivated by this consideration, we
observe that classically we have $E_{6} \subset E_{7} \subset
E_{8}$, while this inclusion is lost when we consider the
corresponding affine algebras (and the same thing is true for the
double and the triple extensions). So we look for an algebra that
may contain all the $E$ series and the $E^{(1)}$ series. This is
possible considering the following non-standard extension of
$E_{8}$ (in which the new simple roots are linked to those of
$E_{8}$ using different fundamental weights). We add the new
simple root:
\begin{equation}
\alpha_{+1}:= -h.r. + k_{+} + l_{+} = - (\varepsilon_{7} +
\varepsilon_{8}) + k_{+} + l_{+}, \quad \alpha_{+1}^{2}=2,
\quad\quad
\end{equation}
 Then we add the two simple roots:
 \be
 \Lambda_{1} = 2 \varepsilon_{8},  \quad\quad \Lambda_{8} = \frac{1}{2} \left( \sum_{i=1}^{7} \, \vr_{i}  + 5 \varepsilon_{8} \right ) ,
 \ee
\begin{equation}
\alpha_{+2}:= - \Lambda_{1} - k_{-} + l_{+} - l_{-}, \quad
\alpha_{+1}^{2}=2,
\end{equation}
and
\begin{equation}
\alpha_{+3}:= - \Lambda_{8} + k_{+} + l_{+} - 3l_{-}, \quad
\alpha_{+3}^{2}=2,
\end{equation}
so that the only non-zero scalar products are:
$(\alpha_{+1},\alpha_{7})=-1$, $(\alpha_{+2},\alpha_{1})=-1$ and
$(\alpha_{+3},\alpha_{8})=-1$ and we have the following Dynkin
diagram:

 
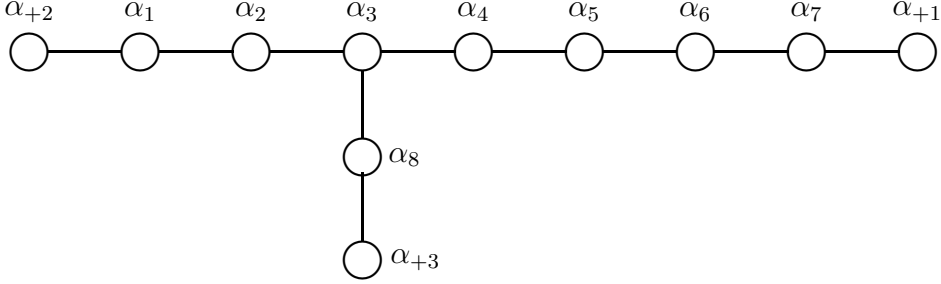
\begin{center}
    \begin{picture}(300,40) \thicklines
        \multiput(-42,20)(42,0){9}{\circle{14}}
        \put(-42,35){\makebox(0,0){$\alpha_{+2}$}}
        \put(0,35){\makebox(0,0){$\alpha_{1}$}}
        \put(42,35){\makebox(0,0){$\alpha_{2}$}}
        \put(84,35){\makebox(0,0){$\alpha_{3}$}}
        \put(126,35){\makebox(0,0){$\alpha_{4}$}}
        \put(168,35){\makebox(0,0){$\alpha_{5}$}}
        \put(210,35){\makebox(0,0){$\alpha_{6}$}}
        \put(252,35){\makebox(0,0){$\alpha_{7}$}}
        \put(294,35){\makebox(0,0){$\alpha_{+1}$}}
        \put(-35,20){\line(1,0){28}}
        \put(7,20){\line(1,0){28}}
        \put(49,20){\line(1,0){28}}
        \put(91,20){\line(1,0){28}}
        \put(133,20){\line(1,0){28}}
        \put(175,20){\line(1,0){28}}
        \put(217,20){\line(1,0){28}}
        \put(259,20){\line(1,0){28}}
        \put(84,13){\line(0,-1){26}}
        \put(84,-26){\line(0,-1){26}}
        \put(84,-59){\circle{14}}
        \put(84,-20){\circle{14}}
        \put(100,-20){\makebox(0,0){$\alpha_{8}$}}
        \put(104,-59){\makebox(0,0){$\alpha_{+3}$}}
    \end{picture}
\end{center}
\bigskip
\bigskip
\bigskip
\bigskip
\bigskip
\bigskip
 \bigskip
This algebra contains $E_{8}$ (then also $E_{7}$ and $E_{6}$) and
all the affinizations $E_{6,7,8}^{(1)}$, so it seems to be
interesting for its content in sub-algebras.
 Let us call $\widehat{\Lambda}_{i} = -  \Lambda_{i}  + k_+  - a k_-$, where $a \in {\mathbb{Z}_{+}}$ and $\Lambda_{i}$,   is any  fundamental weight. We have  ($ \widehat{\Lambda}_{i}^{2} = 2$)
  \bea
& (\widehat{\Lambda}_{1}, \widehat{\Lambda}_{2}) =
(\widehat{\Lambda}_{1}, \widehat{\Lambda}_{5}) = 0  \;\;\;\;
\nonumber \\  (\widehat{\Lambda}_{4}, \widehat{\Lambda}_{8}) = 0 &
(\widehat{\Lambda}_{2}, \widehat{\Lambda}_{3}) =
(\widehat{\Lambda}_{2}, \widehat{\Lambda}_{4}) =
(\widehat{\Lambda}_{2}, \widehat{\Lambda}_{5}) =
(\widehat{\Lambda}_{2}, \widehat{\Lambda}_{6}) = 0
\nonumber \\
&  (\widehat{\Lambda}_{2}, \widehat{\Lambda}_{3}) =
(\widehat{\Lambda}_{2}, \widehat{\Lambda}_{4}) =
(\widehat{\Lambda}_{2}, \widehat{\Lambda}_{5}) =
(\widehat{\Lambda}_{2}, \widehat{\Lambda}_{6}) = 0
 \eea

We have  proposed a procedure to build non standard triple
extended Lie algebras, which we have illustrated with a number of
relevant examples. As already recalled, Kac-Moody or Borcherds
extensions can be defined for these Lie algebras too. On the light
of the remarks of \cite{FN}, it is natural to wonder if these or
some of these algebras are not really subalgebras of the standard
triple extended Lie algebras. This point will be discussed in the
next section, where we discuss also a few examples of subalgebras
which point out the intriguing and surprising structure of the
subalgebras.

\section{Subalgebras of extended Lie algebras} \label{subalgebras}

First of all, let us discuss another non-standard procedure to
extend a Lie algebras of rank $r$. If one adds to the simple roots
of $\mathcal{G}$ the opposite of the h.r.  $\alpha_0 \equiv
\alpha_{r+1}$ and, to recover the linear independence of the
simple root system, one glues the light like vector $k_+$ to a
simple root $\alpha_i$ ($1 \leq i \leq r$), one gets exactly the
affine $\mathcal{G}$. As next step one adds the new root, which
belongs to II$^{1,1}_k$, $\alpha_{r+2}= -(k_+ + k_-)$. From the
Feingold-Nicolai theorem  \cite{FN}, it is easy to realize that in
this way one obtains a generalized Kac-Moody algebra which is
really a subalgebra of the standard overextended Lie algebra
$\mathcal{G}^{++}$.  Things may be different if one considers
simple root systems, obtained by analogous procedure, in the
lattice $\Lambda_{\mathcal{G}} \oplus \mbox{II}^{1,1}_k \oplus
\mbox{II}^{1,1}_l$.  Indeed indefinite Kac-Moody algebras are
obtained which, in general, are described by Dynkin-Kac diagrams
not equivalent to the ones obtained by the standard and not
standard procedure described in the previous section.  A general
discussion of these algebras is beyond the aim of this paper and
we limit ourselves to state a few properties and to present some
examples.  Let us start with the following

{\theorem  The roots $\widehat{\Lambda}_{i} $ defined in Sec.
\ref{nonstandard} are roots of the standard overextended
$\mathcal{G}^{++}$.}

\medskip

Proof:  We shall explicitly write $\widehat{\Lambda}_{i} $  in
terms of the simple roots of $\mathcal{G}^{++}$. Let us remark
that, by construction, $\mathcal{G}^{++}$ contains two affine
$\mathcal{G}$, i.e. $\mathcal{G}^{+}$, whose real root system is
formed by the roots of $\mathcal{G}$ plus $n \, k_{+}$,
respectively $n \, k_{-}$, ($ n \in \mathbf{Z}$). Clearly the
following decomposition in roots is not all unique.
 \begin{enumerate}
 \item $so(2N)$
 \bea
  \widehat{\Lambda}_{2m} & = &  - (- \vr_{1}  - \vr_{2} + k_{+}) + (- \vr_{3}  - \vr_{4} - k_{-}) +
 \ldots + (- \vr_{2m-1}  - \vr_{2m} - k_{-})   \nonumber \\
 & = & - \Lambda_{2m} + k_{+} -(m-1) \,  k_{-}
 \eea
 \item $E_{6}$
 \bea
  \widehat{\Lambda}_{3} & = & \frac{1}{2}[- \sum_{i \neq 6,7} \vr_{i}  + \vr_{6} + \vr_{7}+ k_{+}]  +  \frac{1}{2}[\vr_{1}  + \vr_{2} - \vr_{3}  - \vr_{4} - \vr_{5}  + \vr_{6} + \vr_{7}  - \vr_{8} -2  k_{-}]  \nonumber \\
  & = & - \Lambda_{3} + k_{+} - 2 \,  k_{-}
  \eea
 \item $E_{7}$
 \bea
   \widehat{\Lambda}_{2} & =  & \frac{1}{2}[ \vr_{1}  - \vr_{2} - \vr_{3}  - \vr_{4} - \vr_{5}  - \vr_{6} + \vr_{7}  - \vr_{8} + k_{+}]  + (\vr_{7}   - \vr_{8}   -  2  k_{-})    \nonumber  \\
   & =  & -\Lambda_{2} + k_{+} -  2  k_{-} \\
\widehat{\Lambda}_{3}  & = & (- \vr_{3}   + \vr_{7} + k_{+}) +
(- \vr_{4}   + \vr_{7}  -  2 k_{-})
 +   (- \vr_{5}  - \vr_{8}  -   k_{-})  +   (- \vr_{6}   - \vr_{8}  -  2 k_{-})  \nonumber  \\
 & =  & -\Lambda_{3} + k_{+} -  5  k_{-} \\
  \widehat{\Lambda}_{5}  & = & (- \vr_{3}   - \vr_{6} + k_{+})  +   (- \vr_{7}  + \vr_{8}  -   k_{-})   =
  -\Lambda_{5} + k_{+} -   k_{-}
 \eea
 \item $E_{8}$
 \bea
 \widehat{\Lambda}_{1} &  =   & (- \vr_{i}   -   \vr_{8} + k_{+}) +   (\vr_{i}   -  \vr_{8}   - k_{-})
   =     -\Lambda_{1} + k_{+} -   k_{-}
  \;\;\;\;  (i \neq 8)
 \nonumber \\
   \widehat{\Lambda}_{2} & =  &  \frac{1}{2}[(\vr_{1}  + \vr_{2} + \vr_{3}  + \vr_{4} - \vr_{5}  - \vr_{6} - \vr_{7}  - \vr_{8} )+ k_{+}]  + (-\vr_{3}   - \vr_{8}   -   k_{-})
 \nonumber \\
  & + & (-\vr_{4}  -  \vr_{8}   -  2  k_{-}) + (-\vr_{5}  - \vr_{8}   -   3 k_{-})  =  -\Lambda_{2} + k_{+} -  6  k_{-} \nonumber  \\
    \widehat{\Lambda}_{3}  & = & (- \vr_{3}  -  \vr_{8} + k_{+}) +   (- \vr_{4}   -  \vr_{8}  -  2 k_{-})
 +   (- \vr_{5}  - \vr_{8}  -   3 k_{-})  \nonumber \\
 &+ &  (- \vr_{6}   - \vr_{8}  -  4 k_{-}) +   (- \vr_{7}   - \vr_{8}  -  5  k_{-}) = -\Lambda_{3} + k_{+} -  14  k_{-}
 \nonumber  \\
  \widehat{\Lambda}_{4}  & =  &   (- \vr_{4}   -  \vr_{8}  +  k_{+})
 +   (- \vr_{5}  - \vr_{8}  -   2  k_{-})  +   (- \vr_{6}   - \vr_{8}  -  3 k_{-}) +   (- \vr_{7}   - \vr_{8}  -  4  k_{-})
  \nonumber  \\
  & = &  -\Lambda_{4} + k_{+} -  9  k_{-}
\nonumber \\
 \widehat{\Lambda}_{5}  & = & (- \vr_{5}   - \vr_{8} + k_{+})  +   (- \vr_{6}  - \vr_{8}  -   2 k_{-})
 +  (- \vr_{7}   - \vr_{8}  -  3  k_{-})  \nonumber \\
 & = &  -\Lambda_{5} + k_{+} -  5  k_{-}
\nonumber \\
   \widehat{\Lambda}_{6} & =   & (- \vr_{6}   -   \vr_{8} + k_{+}) +   (-\vr_{7}   -  \vr_{8}   - 2 k_{-})
   =   -\Lambda_{6} + k_{+} -  2  k_{-}
     \nonumber \\
  \widehat{\Lambda}_{8}  & = &  \frac{1}{2}[(-\vr_{1}  - \vr_{2} - \vr_{3}  - \vr_{4} - \vr_{5}  - \vr_{6} + \vr_{7}  + \vr_{8} )+ k_{+}]  + (-\vr_{6}   - \vr_{8}   -   k_{-})   \nonumber \\
 & + & (-\vr_{7}  -  \vr_{8}   -  2  k_{-})   =   -\Lambda_{8} + k_{+} - 3  k_{-}
\nonumber \\\label{eq.e8}
   \eea
   \end{enumerate}
 Therefore,  the considered combinations of the fundamental weights with the vectors of the Lorentzian lattices II$^{1,1}_{k}$ do belong to the root system of the  overextended Lie algebras, even if  the considered fundamental weights do not belong  to the  root system of the native Lie algebra
 \footnote{One can recover the generators by the change  $ \alpha \ra E_{\alpha}$ and $\alpha + \beta \ra  [E_{\alpha}, \, E_{\beta}]$.}.  It follows that the double non standard extensions (called previously  {\textsl{affine extensions}) are all subalgebras, of the same rank, of $\mathcal{G}^{++}$.
Indeed it is easy to verify that the differences $\beta_i  -
\beta_j$  ($\forall i,j;\;\; i \neq j $) of the simple roots
$\beta_i$ of the non standard double extension of $\mathcal{G}$ do
not belong to the root system , therefore satisfying the
conditions of the Feingold-Nicolai theorem.  In particular  we
have

{\theorem The indefinite Kac-Moody algebras of rank  10 described
by the Dynkin-Kac diagrams,  obtained by adding  to the  diagram
of the  affine algebra $E_{8}$, i.e. $E_{9}$, a dot, connected
with a simple  link  to the $j$-th dot of  $E_{9}$ ($1 \leq j \leq
9$),  is a subalgebra of $E_{10}$.}

\medskip
The simple root systems of these subalgebra is formed by the
simple roots of $E_{8}$, by a root $\widehat{\Lambda}_{i}$ given
by eq.(\ref{eq.e8}) and by the root $ -\theta + a k_{-}$ where $a$
is a positive integer such that $ (\widehat{\Lambda}_{i}, \,
-\theta + a k_{-}) = 0$.

One can naturally ask if an analogous theorem holds for  $E_{11}$,
that is if the triple non standard extensions of $E_{8}$ form
Lorentzian algebras of rank 11, subalgebra of  $E_{11}$.   We
have:

{\theorem The indefinite Kac-Moody algebras of rank  11 obtained
by adding  to the   simple root system of the  algebra $E_{8}$
three roots: $ \alpha_{+1} = \widehat{\Lambda}_{j}$,   connected
with a simple  link  to the $j$-th dot of  $E_{8}$ ($1 \leq j \leq
8, \, j \neq 7$);    $ \alpha_{+2} =- \theta - a k_{-}$, where $a$
is a positive integer such that ($ \alpha_{+2}  \, \alpha_{+1} $)
= 0 and  $ \alpha_{+3} = l_{+} + l_{-} - k_{-}$, simply linked
with  $\alpha_{+1} $, is a subalgebra of $E_{11}$.}

\medskip

The proof is straightforward using the explicit expressions of
$\widehat{\Lambda}_{j}$, given in eq.(\ref{eq.e8}). Let us remark
that these subalgebras do not have as subalgebra $E_{10}$. Below
we give the simple roots systems of a set (not exhaustive) of
   rank eleven subalgebra  $E_{11}$, which  contains as subalgebra $E_{10}$.
Let us consider the following simple root system of $E_{11}$:
 $\alpha_i$ ($1 \leq i \leq 8$) are the simple roots of $E_8$, $\alpha_9 = - \alpha_0 - k_+$, $\alpha_{10} =  k_+  +  k_-$ and $\alpha_{11} =  (l_+ +  l_-) - k_+$.
 We follow the convention of  \cite{FSS} and, for the reader convenience, we explicitly  write here the $E_8$  simple root system and the  root system $\Delta$:
\bea & \alpha_1 = \frac{1}{2} \, ( \vr_1 + \vr_8 - \sum_{j=2}^{7}
\,\vr_j)  \;\;\;\;\;\; \alpha_i =   \vr_i - \vr_{i-1}
\;\;\; i = 2,\ldots,7 \nonumber \\
&   \alpha_8 =   \vr_1 + \vr_2  \;\;\;\;\;\;  h.r. : = \alpha_0  =
\vr_7 + \vr_8 \eea \be \Delta   = \left\{ \frac{1}{2} \, ( \pm
\vr_1 \pm \vr_2 \pm  \vr_3 \pm \vr_4 \pm  \vr_5  \pm \vr_6 \pm
\vr_7 \pm \vr_8), \;   \pm \vr_i \pm  \vr_j \right \}
  \ee
  where the total number of $+$ signs (or $-$ signs) in the first expression is an even number.
Let  us consider the $E_{10}$
 Dynkin-Kac diagram obtained by the $E_{11}$ diagram, deleting the dot corresponding to $11$-th  simple root. Let us denote by  $E_{10}^{(j)}$ the algebra of rank 11 whose Dynkin-Kac diagram is
 obtained by the $E_{10}$ diagram adding a dot (in the following denoted by +1) with a simple link to the $j$-th dot ($1 \leq j \leq 10$) and by $\beta_i^{(j)}$ ($ i = 1, \ldots,10, +1$) the simple roots of  $E_{10}^{(j)}$. In the following we do not explicitly write the upper label $j$ in the roots. Clearly $E_{10}^{(10)} = E_{11}$ and, in this case, $\beta_{+1} = \alpha_{11}$ .  We make the following (not unique)  choice for the simple root system of $E_{10}^{(j)}$ ($1 \leq j \leq 9$):
 \begin{itemize}
 \item j = 9)
 \bea
&   \beta_1 = \alpha_1  -  k_+   \;\;\;\;\;   \beta_i = \alpha_i
\;\;\; i = 2, \ldots,7,8
  \nonumber \\
&  \beta_9 = -\vr_{7} + \vr_{8}   +   k_-  \;\;\;\;\; \beta_{10} =
- \alpha_0  -  k_+  \;\;\;\;\; \beta_{+1} = \alpha_{11}
 \eea
 \item j = 7)
 \bea
 &  \beta_i = \alpha_i  \;\;\; i = 1, \ldots, 6, 8 \;\;\;\;\; \beta_7 = \alpha_7  -  k_+
 \nonumber \\
& \beta_9 = \alpha_{10}  \;\;\;\;\; \beta_{10} = \alpha_{11}
\;\;\;\;\; \beta_{+1} = - \alpha_0
 \eea
 \item j = 6)
 \bea
&   \beta_i = \alpha_i  \;\;\; i = 1, \ldots, 5, 7, 8 \;\;\;\;\; \beta_6 = \alpha_6 +  k_-     \nonumber \\
&  \beta_9 =  -\alpha_0  \;\;\;\;\; \beta_{10} =  \vr_{7} +
\vr_{6}  -   k_+  \;\;\;\;\; \beta_{+1} = \alpha_{11}
 \eea
 \item j = 5)
 \bea
&   \beta_i = \alpha_i  \;\;\; i = 1,2,3, 4,6, 7, 8 \;\;\;\;\;
\beta_5 = \alpha_5 +  k_-
      \nonumber \\
&  \beta_9 =  -\alpha_0  \;\;\;\;\; \beta_{10} =  \vr_{5}  +
\vr_{8} -   k_+ \;\;\;\;\; \beta_{+1} = \alpha_{11}
 \eea
  \item j = 4)
 \bea
&   \beta_i = \alpha_i  \;\;\; i = 1, 2, 3, 5,6,7, 8 \;\;\;\;\;
\beta_4 = \alpha_4 +  k_-
   \nonumber \\
&   \beta_9 =  -\alpha_0  \;\;\;\;\; \beta_{10} =  \vr_{4}  +
\vr_{8} -   k_+  \;\;\;\;\; \beta_{+1} = \alpha_{11}
 \eea
 \item j = 3)
 \bea
&  \beta_i =  \alpha_i  \;\;\; i = 1, 2,\ldots,7,8  \;\;\;\;\;
\beta_3 =  \alpha_3  + k_-
    \nonumber \\
& \beta_9 =  -\alpha_0  \;\;\;\;\; \beta_{10} =  \vr_{3}  +
\vr_{8}  -   k_+  \;\;\;\;\; \beta_{+1} = \alpha_{11}
     \eea
 \item j = 2)
 \bea
&  \beta_i =  \alpha_i   \;\;\; i = 1, 3, \ldots,7, 8 \;\;\;\;\;
\beta_2 = \alpha_2 +  k_-
      \nonumber \\
& \beta_9 =  -\alpha_0  \;\;\;\;\; \beta_{10} =  -\vr_{1}  +
\vr_{8}  -   k_+  \;\;\;\;\; \beta_{+1} = \alpha_{11} \eea
  \item j = 1)
 \bea
&   \beta_{1} =   - \frac{1}{2} \,   \sum_{i=i}^{8}  \, \vr_{i}  +
k_- \;\;\;\;\;  \beta_2 = \alpha_8
\;\;\;\;\;  \beta_i =  \alpha_i  \;\;\; i = 3,\ldots,6  \;\;\;\;\; \beta_8 = \alpha_2   \nonumber \\
&  \beta_7 =   \alpha_7 \;\;\;\;\;    \beta_9 =  - \vr_{7}  +
\vr_{8} \;\;\;\;\;  \beta_{10} = - \alpha_1 +   k_+  \;\;\;\;\;
\beta_{+1} =  \alpha_{11}
 \eea
   \item  j = 8)
 \bea
& \beta_i = \alpha_i  \;\;\; i = 1,\ldots,7 \;\;\;\;\; \beta_8 =
\alpha_8 +   k_-
 \nonumber \\
 &  \beta_9 =  -\alpha_0  \;\;\;\;\; \beta_{10} = \frac{1}{2} \, \sum_{i=1}^{8}  \, \vr_{i} -   k_+  \;\;\;\;\; \beta_{+1} = \alpha_{11}
 \eea
\end{itemize}
It is easy to verify that the roots $\beta_i$ belong to the roots
sistems of $E_{11}$,\footnote{Actually they belong, except for
$\alpha_{11} $, to the root system of affine $E_{8} = E_{9}$ which
is a regular sub-algebra of  $E_{11}$} while the differences
$\beta_i  - \beta_j$ ($\forall i,j; \;i \neq j $) ($|\beta_i  -
\beta_j|^{2} \geq 4$) do not belong, therefore satisfying the
conditions of the Feingold-Nicolai theorem.  The Dynkin-Kac
diagrams describing these algebras  contain {\textsl loops} except
for $j = 7$. This algebra  has been considered in \cite{MM}, where
it has been denoted $EE_{11}$, and it has been shown to be a
subalgebra of $E_{11}$ by explicitly constructing the generators
by commutation of the $E_{11}$ generators. The Dynkin-Kac diagrams
for these algebras are easily drawn by adding to the $E_{10}$
diagram a dot simply connected with the $j$-th dot  and then
connecting, with a simple link, the following dots: $ j = 9) \;
\;7-10;  \; j=6) \; \;1-10;   \; j=5) \;  \; 6-10;  \; j=4) \; \;
5-10;  \; j=3) \; \;  4-10;  \; j=2) \; \; 8-10;  \;  j=1) \; \;
7-10, 8-10;  \; j=8) \; \; 1-10$. Of  course, these subalgebras do
not exhaust the set of 11 dimensional indefinite Kac-Moody
subalgebras of $E_{11}$.
In particular we have not considered the subalgebras which do appear as invariant algebras with respect to an involution of the generators of $E_{11}$, see \cite{MM} and \cite{MM05}.

 \section{Conclusions and future developments}

In studying 4-extended Lie algebras, we have seen that Borcherds
algebras seem to emerge naturally. This remark rises the question:
which are the fingerprints of a theory which exhibits a symmetry
under a Borcherds algebra? This question is indeed interesting on
the light of the remark that many dualities have a
group-theoretical origin  in the Weyl group of the algebra. The
Weyl group of the Borcherds algebra has peculiar properties as the
reflection with respect to the imaginary vanishing roots is not
defined. Some particular properties related to this kind of
algebras have already been discussed in \cite{Sla}. The
non-standard extension introduced in this paper have peculiar
features, which deserve further investigation,  on both  their
mathematical structure and their possible physical relevance.  A
classification of these algebras is beyond the aim of this paper,
where we present only a few representative examples. As, however,
very little is known on Lorentzian Kac-Moody algebras, we believe
that any new information is interesting. In the cited literature
on the physical role of the very extended Lie algebras, non-linear
realizations of the indefinite Kac-Moody algebras are used. How a
Chevalley realization of this algebra looks like? In \cite{MS} a
procedure to build up vertex realization of Lorentzian algebra
with only a lattice II$^{1,1}$ has been proposed and applied to
the very simple case of the overextended $A_{1}$ algebra.   It
seems possible to generalize that procedure to the triple extended
Lie algebras. Moreover, it has also been argued by P. West
\cite{West03} that $sl(32)$ is contained in the Cartan invariant
sub-algebra of  $E_{11}$. At first sight the rank of $sl(32)$ is
too large to be a sub-algebra, so it seems that very extended
algebras, at least in the non linear realization, admit finite
dimensional sub-algebras which naively could not be there. The
investigation of the finite Lie subalgebras of the indefinite
Kac-Moody algebras requires new methods beyond the very familiar
ones used in the case of finite Lie algebra, which are essentially
based on the Dynkin methods. This feature is not completely
unrelated with the property, noted in \cite{FN}, that the set of
infinite dimensional sub-algebras of Lorentzian algebras is quite
rich and surprising.  We have illustrated this feature in Sec. 5,
discussing a class of subalgebras of $E_{11}$,  but it would be
useful to dispose of techniques to build up explicitly or to
identify classes of these sub-algebras or to dispose of further
examples.

\medskip

{\bf Acknowledgment} - We thank C. Helfgott, A. Kleinschmidt and
I. Schnakenburg for pointing us a mistake in the previous version
and for useful comments.

\appendix

\section{Some facts about the lattice II$^{1,1}$} \label{lattice}

We review some basic facts about the lattice II$^{1,1}$, which is
the only Lorentzian even self-dual lattice in dimension 2. The
points in this lattice can be described as the vectors:
\begin{equation}
(n,m)
\end{equation}
with $n, m \in \mathbb{\mathbb{Z}}$ and Gram matrix $\left(
\begin{array}{cc}
  0 & -1 \\
  -1 & 0 \\
\end{array}
\right)$, with eigenvalues $\pm 1$. In this way, the scalar
product between two vectors $a=(a_{+},a_{-})$ and
$b=(b_{+},b_{-})$ can be written as: $a \cdot b = -\, a_{+}b_{-} -
a_{-}b_{+}$. We can take $k_{+} \equiv (1,0)$ and $k_{-} \equiv
(0,-1)$ as basis vectors in II$^{1,1}$; with this choice, we have:
\begin{equation}
k_{\pm} \cdot k_{\pm}=0\, ,\quad k_{\pm} \cdot k_{\mp} = 1.
\end{equation}
All the vectors in II$^{1,1}$ can be written as $v = p\, k_{+} +
q\, k_{-}$ (with $p, q \in \mathbb{Z}$); in particular there are
only two vectors of squared norm 2, $\pm (k_{+} + k_{-})$, but
infinite vectors of positive ($\geq 4$) and negative ($\leq -2$)
squared norm, being $v^{2}=2p\,q$.

\section{Definition of Borcherds Algebras} \label{Borcherds}

The best way to think of a Borcherds algebra is to consider it as
a generalization of a finite-dimensional simple Lie algebra. The
definition is based on the Serre-Chevalley construction of
finite-dimensional algebras; we follow \cite{Bor}.These algebras
always have a symmetric matrix and their structure is very similar
to that of ordinary Kac-Moody algebras, the only major difference
is that generalized Kac-Moody algebras allow the presence of
imaginary simple roots. Let $A=(a_{i,j})$ a $n \times n$ (real)
symmetric matrix satisfying the following properties:

\begin{itemize}
    \item $a_{ii}=2$ or $a_{ii} \leq 0$,
    \item $a_{ij} \leq 0$ if $i \neq j$,
    \item $a_{ij} \in \mathbb{Z}$ if $a_{ii}=2$.
\end{itemize}
Then the Borcherds algebra $\mathcal{G}(A)$ associated with the
Cartan matrix $A$ is the Lie algebra given by the following
generators and relations.

\textit{$3n$ Generators}: $e_{i}, f_{i}$ and $h_{i}$

\textit{Relations}:

\begin{itemize}
    \item $[h_{i},h_{j}]=0,$
    \item $[e_{i},f_{j}]=\delta_{ij} h_{i}$,
    \item $[h_{i},e_{j}]=a_{ij} e_{j}, [h_{i},f_{j}]=-a_{ij}
    e_{j}$,
    \item $e_{ij}:=(\mbox{ad } e_{i})^{1-a_{ij}}\, e_{j}=0$, $f_{ij}:=(\mbox{ad } f_{i})^{1-a_{ij}}\,
    f_{j}=0$ if $a_{ii}=2$ and $i \neq j$,
    \item $e_{ij}:= [e_{i},e_{j}]=0$, $f_{ij}:= [f_{i},f_{j}]=0$
    if $a_{ii} \leq 0$, $a_{jj} \leq 0$ and $a_{ij}=0$.
\end{itemize}

The elements $h_{i}$ form a basis  for an abelian subalgebra of
$\mathcal{G}(A)$, called Cartan subalgebra $\mathcal{H}(A)$; as it
happens for Kac-Moody algebras, $\mathcal{G}(A)$ has the
triangular decomposition:
\begin{equation}
\mathcal{G}(A)= \mathcal{N}_{-} \oplus \mathcal{H}(A) \oplus
\mathcal{N}_{+}
\end{equation}
and has many of the properties of the usual Kac-Moody algebras
(real and imaginary roots, etc.). In particular, in this paper, we
have considered Borcherds algebras with just one imaginary simple
root (with squared norm 0), which we have added by hand.
 A rank 2 Borcherds algebra in the lattice II$^{1,1}$ can be constructed as follows:
  the Cartan matrix is given by

  \begin{equation}
A = \left(\begin{array}{cc}
0 & -1  \\
-1 & 2
\end{array}\right)
\end{equation}

 A possible choice for the simple roots is
\be \alpha_1 = k_+  \,,  \;\;\;\;  \alpha_2= -(k_+   +  k_-)\,,
\ee with Weyl vector $\rho= -  k_+$  (defined by  $(\rho, \,
\alpha_i) =  1/2  \; (\alpha_i, \, \alpha_i) $).

\end{document}